\documentclass{rpvol10x7}

\usepackage{enumerate}

\usepackage[english]{babel}

\begin{document}

%%%%%%%%%%%%%%%%%%%%%%
%\titlepages

%\part{Jeannette Nelson}{} % Optional paper

%\blankpage                         % blank page with no running heads

%\vfill

%\clearpage % for consecutive articles

% sample paper 2 from TeX input

\toctitle{Geometric Supergravity}
\tocauthor{R.D'Auria}
\tocsource{{\it Source} {\bf vol}, page (year)}

\chapter{Geometric Supergravity}

\author{R. D'Auria}
\address{Politecnico di Torino, DISAT Department, \\ Corso Duca degli Abruzzi 24, 10129 Torino, Italy.}

%\documentclass[a4paper,12pt]{article}
%\documentclass[aps,pre,nofootinbib,amsmath,amssymb,amsfonts]{revtex4}
%\usepackage{epsfig}
%\usepackage{graphicx}
%\usepackage{dcolumn}
%\usepackage{bm}
%\usepackage{longtable}
%\usepackage{amssymb}
%\usepackage{amsfonts}
%\usepackage[margin=1.0in]{geometry}
%\usepackage{hyperref}
%%\linespread{2.5}
%\usepackage{amsmath}
%\usepackage{showkeys}
%
%%%%%%%%%%%%%%%%%%%%%%%%%%%%%%%%%%%%%%%%%%%%%%%%%%%%%%%%
%%%%%%%%%%%%%%%           Defs          %%%%%%%%%%%%%%%%
%%%%%%%%%%%%%%%%%%%%%%%%%%%%%%%%%%%%%%%%%%%%%%%%%%%%%%%%
%\def\dd{\textrm{d}}                      %%
%\def\wa{\wedge\ast}                  %%
%\def\R{\mathcal{R}}
%\def\I{\mathcal{I}}
%\def\cN{\mathcal{N}}                  %%
%\def\cNb{\bar{\mathcal{N}}}
%               %%
%\newcommand{\1}{\mbox{$1 \hspace{-1.0mm} {\bf l}$}}  %%
%%%%%%%%%%%%%%%%%%%%%%%%%%%%%%%%%%%%%%%%%%%%%%%%%%%%%%%%
%%%                                   %%
%%%%%%%%%%%%%%%%%%%%%%%%%%%%%%%%%%%%%%%%%%%%%%%%%%%%%%%%
%

%%%%%%%%%%%%%%%%%%%%%%%%%%%%%%%%%%%%%%%%%%%%%%%%%%%%%%%%%%%%%%%%%%%%
%%  Comando per aggiungere la sezione nei numeri delle equazioni. %%
%%%%%%%%%%%%%%%%%%%%%%%%%%%%%%%%%%%%%%%%%%%%%%%%%%%%%%%%%%%%%%%%%%%%
\makeatletter                             %%
\renewcommand\theequation{\arabic{equation}}      %%
%\@addtoreset{equation}{section}                      %%
\makeatother                              %%
%%%%%%%%%%%%%%%%%%%%%%%%%%%%%%%%%%%%%%%%%%%%%%%%%%%%%%%%%%%%%%%%%%%%
%%                                    %%
%%%%%%%%%%%%%%%%%%%%%%%%%%%%%%%%%%%%%%%%%%%%%%%%%%%%%%%%%%%%%%%%%%%%

%\usepackage{slashed}

\newcommand{\ft}[2]{{\textstyle\frac{#1}{#2}}}
\def\Re{\mathop{\rm Re}\nolimits}
\def\Im{\mathop{\rm Mi}\nolimits}
\def\trace{\mathop{\rm Rt}\nolimits}
\def\rme{{\rm e}}
\def\rmi{{\rm i}}
\def\rmd{{\rm d}}

\newcommand{\su}{su}
\newcommand{\SU}{SU}
\newcommand{\SO}{SO}
\newcommand{\U}{U}
\newcommand{\USp}{USp}
\newcommand{\OSp}{OSp}
\newcommand{\Symp}{Sp}
\newcommand{\Sl}{S\ell }
\newcommand{\Gl}{G\ell }

\def\a{\alpha}
\def\b{\beta}
\def\g{\gamma}
\def\G{\Gamma}
\def\d{\delta}
\def\D{\Delta}
\def\e{\epsilon}
\def\ve{\varepsilon}
\def\f{\phi}
\def\vf{\varphi}
\def\F{\Phi}
\def\p{\psi}
\def\P{\Psi}
\def\k{\kappa}
\def\l{\lambda}
\def\L{\Lambda}
\def\m{\mu}
\def\n{\nu}
\def\r{\rho}
\def\s{\sigma}

\def\ii{{\rm i}}

%end van ............................................................................................................................................................

\def\bN{\,\mathbf{N}}
\def\bM{\,\mathbf{M}}
\def\bK{\,\mathbf{K}}
\def\bV{\,\mathbf{V}}
\def\bB{\mathbf{B}}

\def\cD{ {\cal D} }

%%%%  Def's for Wigs %%%%

\def\ac{+\nonumber\\&}
\def\acI{+\right.\nonumber\\& \left.}
\def\acII{+\right.\right.\nonumber\\& \left.\left.}
\def\acIII{+\right.\right.\right.\nonumber\\& \left.\left.\left.}

\hyphenation{fern-mi-on-ic}

%%%%%%%%%%%%%%%%%%%%%%%%%%%%%%%%%%%%%%%%%%%%%%%%%%%%%%%%%%%%%%%%%%%%
%%  Comando per aggiungere la sezione nei numeri delle equazioni. %%
%%%%%%%%%%%%%%%%%%%%%%%%%%%%%%%%%%%%%%%%%%%%%%%%%%%%%%%%%%%%%%%%%%%%
\makeatletter                             %%
\renewcommand\theequation{\thesection.\arabic{equation}}      %%
\@addtoreset{equation}{section}                   %%
\makeatother                              %%

\section{Introduction{.}}

In the year 1978 Y. Ne'eman and {T.} Regge \cite{Neeman:1978njh} proposed a new approach to the formulation of gauge
theories, specifically gravity and supergravity, based {on the} formalism introduced by E. Cartan for the formulation  of Riemannian geometry
in a completely geometrical setting. Cartan's approach  implies a new{,} more geometrical and group-theoretical way of formulating  General Relativity. Indeed, as the adopted formalism relies basically and consistently on the use {of differential} forms, the Cartan's beautiful setting is independent of different coordinate frames, that is of the group of general coordinate transformations (GCTG). At the same time{,} it gives a prominent role to the gauge invariance of the theory under the Lorentz group which emerges quite naturally from the formalism. As a matter of fact, in Cartan's view, Riemannian geometry must be seen as pertaining to finite Lie groups rather than to infinite group of general coordinate transformations. In the latter case it could  be difficult to see how gravitation could be unified with gauge theories of other interactions, what instead seems quite natural in the geometrical formalism developed by him.\\
Following this line of approach{,} Y. Ne'eman and {T.} Regge further developed the Cartan's formalism{,} proposing that in principle any diffeomorphic and  gauge invariant theory should be constructed directly on \emph{the group manifold} G defining the Lie algebra valued gauge fields in the coadjoint representation {of the} group. This is consequence of the following considerations.

 Referring to the pure gravity theory, the spin connection $\omega^{ab}$ and the vierbein $V^a$ {1-form} gauge fields are just fragments of the adjoint multiplet $\mu^A$, $(A=1,\dots,10)$, of the Poncar\'e Lie algebra with indices  decomposed with respect to the Lorentz subgroup $\rm SO(1,3)$, and spanning a basis of the cotangent plane of the Poincare' group,  namely $\mu^A= (\omega^{ab},V^a)$, {$(a,b=0,\dots ,3)$}. However in gravity theory, there is an essential difference between the two fields: While the vierbein $V^a=V^a_\mu\,dx^\mu$ propagate, this is not true for the spin connection. This is a consequence of the well known fact that Riemannian geometry requires that the {``}curvature" of $V^a$, namely the torsion 2-form, must vanish. This makes the spin connection a functional of the vierbein and its derivatives. This disparity is essentially due to the \emph{factorization hypothesis}  which breaks from the very beginning the symmetry of the  group, since it factorizes in a trivial way the dependence of {the} gauge fields from the Lorentz coordinates. In fact in the Cartan formulation of gravity the fields $(\omega^{ab},V^a)$ live on the principal fiber bundle $[\mathcal M_4/\,SO(1,3)]$, where the base space $\mathcal M_4$ is the physical space-time which, in the vacuum configuration, reduces to the coset base space $\rm [G/SO(1,3)]$ with $\rm G=\rm ISO(1,3)$.

 It follows that if we now do not assume factorization, the dependence of the fields $\mu^A= (\omega^{ab},V^a)$ on the group coordinates must be dictated by the field equations (and boundary conditions). It is then  natural to try to construct gravity theory directly on the group manifold $\rm G$,  $(\rm ISO(1,3)$ in the gravity case), where the fields are represented by the Cartan-Maurer 1-forms in the coadjoint representation of the group. This implies that the gauge fields will depend {not only on} the coordinates $x^\mu$ {$(\mu=0,\dots,3)$} related to the translation group, {but also} also on the coordinates of the Lorentz group $y^{\mu\nu}$.  Out of the vacuum the left-invariant 1-forms,  spanning a basis on the cotangent space of $\rm G$, do not obey anymore the Cartan-Maurer equations, but become dynamical, that is they acquire \emph{curvatures}, namely the field-strengths of the  dynamical fields.

Quite generally, as can be shown in general and we will show in the simplest cases, when fields are defined directly on a {(graded-)}group manifold, this new approach makes it possible to give a more geometrical formulation of the theory. Actually, both for  gravity theory as for supergravity in any dimensions $D$, \emph{the Lorentz invariance can be retrieved from the {equations} of motion} as a result of the action principle, even if the $D$-form {Lagrangian} has to be integrated on a group manifold whose dimensionality, $ \rm dim\, G$, is greater than the form degree $D$ of the  {Lagrangian}. In fact we will see that the integration of a $D$-form as a submanifold can be consistently performed as a result of its invariance under {the GCTG}, and the resulting equations of motion give \emph{horizontality} of the curvatures in the Lorentz directions, leading to factorization of the Lorentz parameters. While this way to obtain Lorentz invariance starting from the whole group manifold G seems to be  only of academic interest, its extension to graded groups ({or supergroup,} SG in the following) leads to a {\textit{geometric interpretation of supersymmetry.}} Indeed{,} in this case, referring for simplicity to {the $\mathcal{N}=1$, $D=4$} case, the coadjoint {supermultiplet now} contains an extra fermionic vielbein, namely $\mu^A= (\omega^{ab},V^a, \psi^\alpha)$, {$a,b=0,\dots, 3, \alpha=1,\ldots,4$}, where $\psi^\alpha$ is a Majorana spinor 1-form, so that the coadjoint multiplet  will now depend, besides translation and Lorentz coordinates, also on the odd fermionic Grassman parameters $\theta^\alpha$. In this case the Lorentz factorization, obtained as a result of the field equations, is not sufficient alone to leave us on space-time, but rather on \emph{superspace} defined in the vacuum configuration  as ${\rm R^{4|4}}= \overline{\rm O{S}p(1,4)}/SO(1,3)$, while out of vacuum superspace is a bundle whose base space is physical space-time. The  superspace equations of motion, besides horizontality of the supercurvatures in the Lorentz directions, also give constraints on the super-curvatures in the {``}fermionic" directions, which allow to restrict the theory to space-time only. Indeed one finds that these components are linearly expressible in terms of the components of the super-curvatures on the (cotangent plane of {the}) space-time manifold. It is this property, dubbed \emph{rheonomy}{,} which allows a complete geometrical interpretation of supersymmetry\footnote{The same mechanisms of factorization and rheonomy also work in rigid theories. However, for lack of space we shall not consider theories where rigid supersymmetry is present.} and moreover {allows} the interpretation of the \emph{superspace diffeomorphisms} as supersymmetry {transformations} on space-time.\\

In the following we shall try to give a short, albeit almost complete, account of these properties in the simplest case {of $\mathcal{N}=1$, $D=4$} pure supergravity. This is however sufficient, since, as we will explain, they work exactly in the {same} way for any other supergravity independently of the supergroup $\rm G$ the number of supersymmetry generators, space-time {dimensionality} and/or matter couplings, even if they often exhibit a much more intricate structure. Notwithstanding this the geometrical interpretation of supersymmetry can be shown to remain the same.

The {development} of this  approach was proposed by T. Regge as soon as he came back to Torino from Princeton IAS. The developments of his ideas were initially a result of his collaboration  with R. D'Auria and P. Fr\'e and partly with A. D'Adda. The Torino group then developed the Regge initial ideas during many years\footnote{Torino group was essentially composed by D'Auria and Fr\'e in the beginning, and later enlarged to L. Castellani and many other collaborators  among whom an important role has been played by A. Ceresole. During the development of the approach many other collaborators  joined our group in view of solving specific problems using, at least in part, our techniques, the most assiduous and important being S. Ferrara, L. Andrianopoli and lately M. Trigiante.}  and his geometric method has become one of the principal tools for investigation not only of supergravity, but also of any other topic where the geometrical and group{-}theoretical approach can be useful or even essential. Indeed a long series of  achievements using the basic idea of Tullio Regge has been realized from the very beginning till our days and will probably continue also in the future.
\vskip0.5cm
Coming back to supergravity, let me just mention which  other advantages of a purely geometric approach have emerged during the developments of the method.
\begin{itemize}
\item
An important step has been the construction of the {Lagrangian}. Indeed, using the building principles  of \emph{geometricity} to be discussed below together with other obvious requirements like the presence of the Einstein term{,} the construction turns out to be essentially  algorithmic and unique.Moreover, as outlined before{,} using the Ne'eman Regge's geometrical action principle the superspace equations of motion give in one stroke, besides the space-time field equations, also linear relations between the components of the supercurvatures leading to the supersymmetry of the space-time {Lagrangian}.
\item
Besides, the steady use of {the} geometric approach also for theories having antisymmetric tensors in the gravitational multiplet has led the authors of reference \cite{D'Auria:1982nx} to develop in a geometrical way a new structure for their treatment, by generalizing the Maurer-Cartan equations to integrable structures containing higher degree p-forms.\footnote{The new structures generalizing Maurer-Cartan equations have led to a formulation of a mathematical structure which is nowadays recognized as a first example of the mathematical  theory of $L_\infty$ algebras (see e.g. references \cite{Sati:2015yda, stronghom}).}
\item
As is well known, it is possible to develop supergravity theories in superspace  using Bianchi identities. Since the geometric approach is essentially a superspace approach  it is possible to derive directly (that is without writing an action principle) the equations of motion of any supergravity using only the Bianchi identities of the super-curvatures and using a priori $rheonomy$ as a principle of the construction, as we shall explain in the following.
\end{itemize}

\section{Pure Gravity in Cartan {F}ormalism.}

In this section we shall first remind some of the most important properties of the Cartan formulation of the Einstein gravity in order to establish the notations and thus setting the stage for the formulation of its extension to the Poincar\'e group manifold. This is a preparatory discussion in view of discussing  the geometrical interpretation of supersymmetry (also called \emph{rheonomy}) in supergravity theories.

  In the Cartan geometrical framework the gauge fields are to be identified with the left-invariant differential forms $\{\sigma^A\}$ dual to the generators $T_A$ of the Lie algebra of the Poincar\'e group G= \rm ISO(1,3), namely we have $\sigma^A(T_B)=\delta^A_B$. The indices $A,B$ {run} on the (co{-})adjoint representation of a Lie algebra. The left-invariant 1-forms satisfy the celebrated Maurer-Cartan equations which are a dual formulation of the Lie algebra:\footnote{Here and in the following we shall omit the wedge symbol for $p$-forms product, unless where it can be useful for clarity and avoid confusion.}
\begin{equation}\label{Cartan}
d \sigma^A +\frac12 C^A_{\;BC}  \sigma^B\sigma^C=0.
\end{equation}
Here $ C^A_{\;BC}$ are the structure constants of the Poincar\'e Lie algebra satisfying the Jacobi identities as a consequence the  integrability condition $d^2=0$ of Eq. (\ref{Cartan}).\\
The left-invariant 1-forms $\sigma^A $ describe the  configuration of the physical vacuum, that is,  vanishing field{-}strengths. In order to have non{-}vanishing field{-}strengths one needs a non{-}vanishing right hand side in equation (\ref{Cartan}), that is we must endow G with a set of \emph{non left-invariant forms} $\mu^A$, {so that they can develop non-vanishing curvatures $R^A$:}
\begin{equation}\label{nonleft}
 R^A=   d \mu^A +\frac12 C^A_{\;BC}  \mu^B\mu^C,
\end{equation}
$R^A$ being defined as the coadjoint multiplet of curvatures.
In particular{,} the 1-forms $\mu^A$ will be now dual to the non left-invariant vector fields $\tilde T_A$ closing a Lie algebra of vector fields with structure functions instead of structure constants, namely $C^A_{\;BC}\longrightarrow C^A_{\;BC}+R^A_{\;BC}$, where  $R^A_{\;BC}$ are the components of the curvature 2{-}forms along $\mu^B\mu^C$.
Following Cartan's geometrical setting{,} the 1-forms $\mu^A$, indexed in the coadjoint representation of the Poincar\'e group, can be decomposed {into} their Lorentz content, that is  the index $A$ is decomposed with respect to indices of the  Lorentz subgroup $\rm SO(1,3)$. In this way one defines the spin connection and the vierbein  1-forms: $\mu^A\equiv\{\omega^a_{\,\,b},\,V^a\}$, (${a,b = 0,1,2,3}$). Correspondingly{,} the curvatures or field-strengths of $\omega^a_{\,\,b},\,V^a$ take the following form:
\begin{eqnarray}
  \label{curv}  R^a_{\;\;b} &=& d\omega^a_{\,\,\,b}- \omega^a_{\,\,\,c} \wedge\omega^c_{\,\,\,b}{,}  \\
  \label{tors} T^a &=& dV^a-\omega^a_{\,\,\,b} \wedge V^b\,.
  \end{eqnarray}
The $R^{ab}$ 2-form is named the Lorentz curvature and when expanded along a vierbein basis $R^{ab}=\frac12 R^{ab}_{{\,\,\,cd}}\,V^c\,V^d$ {its} components coincide with minus the Riemann tensor in Lorentz indices{.}\footnote{We note that our definition of curvature and torsion differ by the sign of the spin connection $\omega ^{ab}$ compared to the more common definition in the literature. Since $\omega ^{ab}\rightarrow -\omega ^{ab}$ implies $R^{ab}\rightarrow -R^{ab}$, the relation between $R^{ab}$ and the Riemann tensor is $R^{ab}_{\,\,\,cd}= - V^a_\mu\,V^{b\nu}\,V^\rho_c\, V^\sigma_d\,\,R^{\mu}_{\nu\rho\sigma}$.} The $T^a$ 2-form is named the torsion.
 Moreover{,} the curvatures satisfy the Bianchi identities
\begin{eqnarray}
\label{bianchis} \mathcal D R^{ab}&=&0 {,}  \\
\label{bianchis1} \mathcal D T^a-R^{ab}\,V_b  &=&0 {,}
\end{eqnarray}
{being $\mathcal{D} = d- \omega$ the Lorentz covariant derivative.}
 Using these definitions{,} the Cartan gravitational {Lagrangian} can be written as follows:
\begin{equation}\label{lagr}
  \mathcal L\simeq R^{ab}\wedge V^c\wedge V^d {\epsilon}_{abcd}\,.
\end{equation}
Note that the integrand of the action being  a 4-form  is automatically invariant under diffeomorphisms and, as it is shown below, it \emph{formally} coincides with the usual Einstein-Hilbert {Lagrangian} written in terms of the metric and Levi-Civita connection.
%\footnote{Note that our definition of curvature and torsion has a different sign in the spin connection compared with the definition mostly used in the literature. This implies in particular that the Lorentz curvature is related to Riemann tensor by a minus sign.}

In the original Cartan construction of the Lagrangian (\ref{lagr})  the gauge fields $\{\omega^a_{\,\,b},\,V^a\}$  depend only on the space-time coordinates  while the dependence on the Lorentz parameters is factorized. In other words{,} the total space is taken to be a principal fiber bundle  $[{\mathcal M}_4, \rm{H}]${, where $\rm {H}=\rm {SO}(1,3)$.}

As the fiber bundle structure implies a factorization of the Lorentz parameters of the fiber, ${\mathcal M}_4$ can be identified with a, {generally} non-flat, four manifold  namely  \emph{space-time}. Therefore the gravitational action is obtained by integrating on  $M_4$  the {Lagrangian}  (\ref{lagr}):
\begin{equation}\label{action}
    \mathcal A= \frac{1}{4 \kappa^2}\int_{\rm M_4} R^{ab}\wedge V^c\wedge V^d {\epsilon}_{abcd}\,{,}
\end{equation}
where $\kappa=\sqrt{8\pi\,G}$, and  G is the gravitational constant.

Let us remind  some of the properties of the Cartan {Lagrangian} (\ref{lagr}).\\
First we show that it is \emph{formally} equivalent to the traditional Einstein-Hilbert {Lagrangian}. Indeed\footnote{Here and in the following we are using a mostly minus Minkowski metric.}
\begin{eqnarray}
	&&R^{ab} {\wedge} V^{c}\wedge V^{d} \epsilon_{abcd}= \frac12 R^{ab}_{\phantom{ab}ij}V^{i} V^{j} V^{c} V^{d}\epsilon_{abcd} =
  \frac12 R^{ab}_{\phantom{ab}ij}V^{i}_{\phantom{i}\mu} V^{j}_{\phantom{j}\nu}
	 V^{c}_{\phantom{j}\rho} V^{d}_{\phantom{j}\sigma}
	\epsilon_{abcd}
	d x^{\mu} d x^{\nu} d x^{\rho}d x^{\sigma}= \nonumber \\
	 &&\frac12 R^{ab}_{\phantom{ab}ij}V^{i}_{\phantom{i}\mu} V^{j}_{\phantom{j}\nu}
	 V^{c}_{\phantom{j}\rho} V^{d}_{\phantom{j}\sigma}
	 \epsilon^{\mu\nu\rho\sigma}\epsilon_{abcd}d^{4}x=
	 \frac12 R^{ab}_{\phantom{ab}ij}\epsilon^{ijcd}\epsilon_{{ab}cd}\textrm{det} Vd^{4}x
	=-2 R^{ij}_{\phantom{ab}ij} \textrm{det}V d^{4}x
	\ .\label{equiv}
\end{eqnarray}
If we denote  world-indices  by Greek letters, we have
\begin{equation}
	R^{ij}_{\;\;ij}
	\equiv R^{\mu\nu}_{\;\;\mu\nu}
	= R {,}
	\label{I.4.4}
\end{equation}
where $R$ is the scalar curvature and $\textrm{det}V=\sqrt{-g}$ is the square root of the metric determinant ($g=\textrm{det}g_{\mu\nu}$). Hence we get:
\begin{equation}
	\int_{M_{4}} R^{ab}\wedge V^{c}\wedge V^{d}\epsilon_{abcd} = -2 \int_{M_{4}} R\sqrt{-g}\,d x^{4}
	\ .
	\label{I.4.5}
\end{equation}
Let us now observe that the formal equivalence between the Cartan and Einstein-Hilbert formulations just shown does not mean that they are completely equivalent.\\
First of all Cartan vierbein formalism, showing explicitly the gauge invariance of the theory under Lorentz transformations, allows to introduce spinors in the {G}eneral {R}elativity framework, contrary to what happens in the usual formalism. Indeed in the world index setting, tensors  transform under {$\rm {GL}(4,\rm R)$} while spinors are in a $\rm SO(1,3)$ representation and therefore they can be naturally coupled in a formalism where Lorentz $\rm SO(1,3)$ covariance is present.\\
Furthermore, it is a \emph{first order {Lagrangian}}, that is the gauge fields $\omega^{ab},V^a$, being member of the same adjoint multiplet, are off-shell independent as it is natural in a {geometric Lagrangian} like (\ref{lagr}).  By \emph{geometric} we mean that, besides containing the Einstein term, it is built only in terms of forms and wedge products.\\ Moreover it can be easily ascertained that requiring the {Lagrangian} to be \emph{geometric} makes it \emph{uniquely determined}.
 Indeed any other  Lorentz invariant 4-forms to be added to Eq. (\ref{lagr}) would be a wedge product of curvature 2-forms and would give rise to Chern characteristic classes\footnote{By curvature we mean both the $R^{ab}$ and $T^a$ 2-forms.} (note that the term $T^a\wedge T_a$ is easily seen to reduce to the Cartan Lagrangian by partial integration).\\ Alternatively,  uniqueness of the gravity Lagrangian is also obtained by   requiring that no dimensional constant should enter the {Lagrangian}. In this case the (wedge) product of any two curvatures $R^A=(R^{ab},T^a)$ would have different scaling with respect to the {Lagrangian} (\ref{lagr}) and should therefore be omitted. Of course dimensional constants should appear in theories where coupling to matter, gauging and scalar potential appear.\footnote{
For pure theories, however, like those described in terms of massless fields only, the pure gravity case being the simplest case, one  dimensional constant  of dimensions mass squared is allowed, adding the term $1/3 \Lambda\epsilon_{abcd} V^a\,V^b\,V^c\,V^d $ with $\Lambda$ having the dimension of a mass squared. This gives rise to a Einstein {Lagrangian} with a cosmological term. This kind of extensions, however, can be easily shown to be equivalent to starting with the group manifold of a (anti) de Sitter group instead of the Poincar\'e group and will not change anything in the  mechanisms we are going to discuss both for gravity  as for supergravity. Indeed we may note that the Poincar\'e group $\rm ISO(1,3)$ is an {In\"{o}n\"{u}}-Wigner contraction of the $\rm SO(2,3)$ group.}

{Let us now write down the equations of motion  derived from the action (\ref{action}). Varying the action with respect to $\omega ^{ab}$ and $V^d$ we find, respectively:}
{\begin{eqnarray}
 \label{deltaomega} T^c \wedge V^d \epsilon_{abcd} &=& 0 , \\
  \label{deltavier} R^{ab}\wedge V^c \epsilon_{abcd} &=& 0 ,
\end{eqnarray}
where $T^a= \mathcal DV^a$ is the torsion 2-form,
 ${\mathcal DV^a \equiv dV^a-\omega^a_{\;\;b}  V^b}$ denoting the Lorentz covariant derivative.}

 {Before proceeding with the solutions to the above equations,} it is important to stress that, besides the obvious invariance under GCTG, even if all the fields are valued in  the Poincar\'e group, \emph{the action is invariant only with respect to the subgroup of the (local) Lorentz transformations}.\\
 This can be easily checked  writing down the infinitesimal action of the Poincar\'e group on the gauge fields $\omega^{ab},V^a$. {Defining} $\epsilon^A=\epsilon^{ab},\epsilon^{a}${, being $\epsilon^{ab}$ and $\epsilon^{a}$} the parameters of the infinitesimal Lorentz and translation gauge transformations, respectively, we have:
\begin{equation}\label{adj1}
   \delta^{(gauge)} \mu^{{A}} = \left(\nabla \epsilon\right)^A {,}
\end{equation}
 where $\nabla$  denotes the Poincar\'e gauge covariant differential. Decomposing the (co)-adjoint index $A$ in indices of the Lorentz subgroup, from (\ref{adj1}) follows
\begin{eqnarray}
 \delta^{(gauge)} \omega^{ab} &=& \mathcal D\epsilon^{ab} {,} \nonumber \\
\delta^{(gauge)} V^a &=& \mathcal D\epsilon^a + \epsilon ^{ab} V_b {,} \nonumber
\end{eqnarray}
where $\mathcal D= d-\omega$ denotes the Lorentz covariant differential.
It is then easy to see that the Lagrangian (\ref{lagr}) and the equations of motion are  invariant under a local  Lorentz transformation, but are not invariant under a local translation. Indeed performing an infinitesimal translation on (\ref{action1}), using  {the} Bianchi identities (\ref{bianchis}) and (\ref{bianchis1}) and integrating by parts, we have
\begin{equation}\label{trans}
 \delta \int R^{ab}V^cV^d  {\epsilon_{abcd}} =2 \int R^{ab}\mathcal D \epsilon^{c}V^d {\epsilon_{abcd}} =-2 \int \epsilon^c R^{ab} T^d {\epsilon_{abcd}} \neq 0\,.
\end{equation}
 The non-invariance under translations of the equations of motion (\ref{deltaomega}) {and} (\ref{deltavier}) {can be} checked in an analogous way. We will see {in the next section} that the fact that a torsionless vierbein can acquire torsion under the action of a translation can be {best understood using} the notion of Lie derivative.

 Let us now solve the equations of motion {(\ref{deltaomega})} and (\ref{deltavier}). From (\ref{deltaomega}), expanding the torsion 2-form along the vierbein basis, $ T^a ={T^a_{\phantom{a}bc}} V^b\,V^c ${,} it is easy to find that the components ${T^a_{\phantom{a}bc}}$ have a vanishing  trace,  ${T^a_{\phantom{a}ab}}=0$. This  implies in turn ${T^a_{\phantom{a}bc}} =0 $ as the  unique solution. Therefore
\begin{equation}\label{tor}
  T^a=0 \longrightarrow dV^a- {\omega^a_{\;\;b} V^b=0 .}
\end{equation}
Expanding along the differentials{, we write}
\begin{equation}\label{exp}
  \partial_{[\mu} V^a_{\nu]}=\omega ^{ab}_{[\mu}V_{\nu]b} {.}
\end{equation}
 {Equation} (\ref{exp}) can be solved in an analogous way as the Levi-Civita connection is solved in terms of the metric and its derivatives, obtaining
$\omega^{ab}_\mu$ in terms of the vierbein components and its derivatives.
 Therefore, implementing the purely algebraic equation of motion  of the spin connection allows us to express $\omega^a_{\,\,b|\mu}$ as a functional of the vierbein and its first derivatives. This is strictly analogous to what happens in the first order Palatini formalism in the usual  metric approach.

 From the second equation, expanding the 2-form $R_{ab}= \frac12 R^{ab}_{\;\;\;cd}V^cV^d$, we also find, after some algebraic manipulation, the Einstein equation \footnote{Note that we write  Einstein equation $R_{\mu\nu} -1/2 g_{\mu\nu} R=0$ using rigid vierbein indices. Indeed,  unless necessary, we are using rigid tangent indices throughout instead of coordinate indices. Of course the  rigid ndices  can be traded with the world indices using the vierbein matrix in the usual way.}
\begin{equation}\label{einstein}
  R_{ab}- \frac12\eta_{ab} R=0 {,}
\end{equation}
where now $ R_{ab}$ is the Ricci tensor $R_{ab}= R^{am}_{\;\;\;bm}$ and $R\equiv R^a_{\;\;\;a}$ is the scalar curvature.

\subsection{Extending the Theory from {$\rm G/\rm H$} to $\rm G$.}

We have seen in the previous section that the Einstein-Cartan Lagrangian is  invariant under gauge Lorentz gauge transformations. It is therefore convenient to use the language of fiber bundles.\\ We will refer in the following to the Poincar\'e group, but most of the considerations are exactly the same for other gravity or supergravity theories, replacing  the Poincare' group with a general (super-)group G and the Lorentz group with a general gauge group H.\\
We know that in the vacuum configuration, the fiber bundle is  $[\rm ISO(1,3)/\rm SO(1,3)]$, while for a general configuration the gauge fields $\mu^A =\left(\omega^{ab},\,V^a\right)$ of the Cartan-Einstein Lagrangian (\ref{lagr}) live on a fiber bundle $[\rm  M_4,\rm H]$, where $\rm M_4$ is the base space and   $\rm H =\rm SO(1,3)$ is the fiber.
It is then  convenient and natural to consider the base space $\mathcal M_4$ as a deformation of the vacuum base space $\rm ISO(1,3)$. Indeed the base space out the vacuum
 can be obtained when  the left-invariant 1-forms are deformed into a \emph{non-left-invariant} dynamical forms $\mu^A$ enjoying \emph{curvatures}. In this case the base space can be thought of as a  space $\rm  G/H$ on which a dynamical metric has been defined, constructed out of the on left-invariant 1-forms $\mu^A$, which no longer has G as isometry group. Equivalently we may also say that the group $\rm G$ has been deformed into a $\rm \tilde G$ so has to have  dynamical  fields. $\rm \tilde G$ is often referred to as a \emph{soft group manifold}. With this nomenclature we can write the general structure of the fiber bundle as $\rm\tilde G/\rm H$.

Let us now take the point of view of reference  \cite{Neeman:1978njh} and assume that the set of 1-forms $\mu^A=(\omega^{ab},\,V^a)$  is  defined, right from the beginning,  on the whole Poincar\'e group G=$\rm ISO(1,3)$. In this case factorization of the Lorentz coordinates is absent since the gauge fields $\mu^A=\left(\omega^{ ab},\,V^a\right)$ will now depend on the full set of ten{-}dimensional group coordinates, namely the $x^\mu$  parameters of translations and the coordinates $y^{\mu\nu}$ associated with the Lorentz transformations of $\rm SO(1,3)$.  The non left-invariant $\mu^A$  1-forms, generalizing the $\sigma^A${,} are  naturally viewed as a set of vielbein spanning a local reference frame on the ten{-}dimensional cotangent plane at each point of $\rm \tilde G/\rm H$.

We consider the {Lagrangian} $\mathcal L=R^{ab}\wedge V^{c}\wedge V^{d}\epsilon_{abcd}$ where now both the vielbein and curvatures are not factorized, but depend of the full set of the {$\rm G$}-coordinates $(x^\mu,\,y^{\mu\nu})$.

Even if the \emph{{Lagrangian}} (\ref{lagr}) is formally the same when the fields 1-form  and curvatures 2-forms are defined on the full group manifold of  G, there is, however, a  problem to write down a suitable \emph{action}, since we have to integrate the {Lagrangian} 4-form  in the $d$-dimensional space of the group {$\rm\tilde G$}, ($d=10$ in the Poincar\'e case).\\  A simple way out would be to embed the four-dimensional space-time as a four{-}dimensional hypersurface $\mathcal M_4$ (with boundary value $\rm M_4$) in {$\rm\tilde G$}. However, the very presence of $\mathcal M_4$ in the variational principle makes it a dynamical variable, thus subjected to variation. Indeed new fields enter in the action corresponding to the embedding functions of $\mathcal M_4$ in $\rm \tilde G$. This implies the added complication that the equations describing the embedding of
$\mathcal M_4$ in {$\rm\tilde G$} should include arbitrary functions which must be  considered as fields, what  would of course spoil the geometric nature of the Cartan geometric approach.\\
The crucial observation  given in \cite{Neeman:1978njh} is that one can safely ignore the variation of $\mathcal M_4$, since \emph{any variation of $\mathcal M_4$ can be compensated by a change of coordinates $x^M=(x^\mu,\,\eta^{\mu\nu})$ in {$\rm\tilde G$}} under which the {Lagrangian}, built only in terms {of differential forms (and wedge products among them)}, is invariant.
 When considering theories more general than pure gravity, this of course requires that the {Lagrangian} be \emph{geometrical} in a larger sense as was previously explained. Indeed, besides the requirements of being built using only differential forms, wedge products and the differential operator $d:d^2=0$, we must also add the requirement of \emph{excluding the presence of the Hodge duality operator}. Indeed as the hypersurface $\mathcal M_4$ on which the integration is performed can be chosen arbitrarily,   due to invariance under diffeomorphisms, the equations of motion will hold on the whole $\rm\tilde G$. The presence  of the Hodge duality operator, instead,  would give a dependence  on the hypersurface $\mathcal M_4$ and its metric.

In the case under study, namely simple gravity in the Cartan formalism, see equation (\ref{lagr}), the requirement of geometricity in this larger sense {is} obviously satisfied and  the action will be now written as follows:
\begin{equation}\label{action1}
    \mathcal A= \frac{1}{4 \kappa^2}\int_{\mathcal M_4\subset\rm\tilde G} R^{ab}\wedge V^c\wedge V^d \epsilon_{abcd},
\end{equation}
where $\mathcal M_4$  is an arbitrary surface immersed in $\rm\tilde G=\rm\tilde ISO(1,3)$.\\
 The equations of motion derived from the action (\ref{action1}) are \emph{formally} the same  as  equations (\ref{deltaomega}) and (\ref{deltavier}), but are now valid on the whole group manifold, since{,} as  already observed, the hypersurface $\mathcal M_4$ is arbitrary. Because of that, in analyzing their content, the Lorentz curvature and the torsion 2-forms must be expanded not only in terms of the vierbein  $V^a \wedge V^b$, but also in terms of {the 2-forms} $\omega^{ab}\wedge V^c$ and $\omega^{ab}\wedge \omega^{cd}$, which  are part of the $\mu^{{A}}\wedge \mu^B$ basis of the cotangent plane to {$\rm\tilde G$} at each point of $\mathcal M_4$. The projection along $V^a \wedge V^b$  {leads} of course to the same equations  as in the Cartan original setting. On the other hand, expanding the curvature and the torsion equations also along the other two 2-forms of the basis containing at least one $\omega^{ab}$  1-form, it is almost immediate to recognize that we obtain for their corresponding components the solution:
\begin{equation}\label{proj}
  R^{ab}_{c|(lm)}= R^{ab}_{(lm)|(pq)}=T^a_{c|(lm)}= T^a_{(lm)|(pq)}=0.
\end{equation}
These equations assert that the  curvatures are \emph{horizontal} along the Lorentz subgroup 1-forms containing at least one Lorentz gauge field $\omega^{ab}$, implying that \emph{the group manifold {$\rm\tilde G$} has acquired dynamically the structure of a fiber bundle, namely $[\rm\tilde G/ {\rm H} ,\rm H]$}, where in the present case {$\rm\tilde G=\tilde ISO(1,3), \, \rm H=SO(1,3)$}.
Indeed we recall that an equivalent way to say that a manifold  $\rm\tilde G$ has the structure of a fiber bundle $[\rm\tilde G/ \rm H,\rm H]$ is to require  that the curvatures  $R^A$  are \emph{horizontal}, that is that their components along the $\rm H$ directions spanned by the 1-forms dual to the generators of $H$ vanish. This is in fact the content of  equation (\ref{proj}) since  the {curvature 2-forms} $\left ( R^{ab}\,, T^a\right)$ vanish along components  $\omega^{ab}\wedge \omega^{cd}$ or $\omega^{ab}\wedge V^c$. It follows that only the components along two vielbein $R^A=R^A_{ab} V^a\wedge V^b$ will survive, that is, identifying $\mathcal M_4$ with the space-time and projecting them along the differentials, namely the  $R^A_{\mu\nu}$  components.\\
Thus we have reobtained from the same Lagrangian, but an enlarged  action principle, the same equations of motion as in the classical treatment of the Cartan {Lagrangian} given before.
The new procedure of  defining fields on the group manifold and of considering  space-time as a hypersurface immersed in G, gives a conceptual advantage with respect to the usual formulation since the factorization of the Lorentz coordinates and the Lorentz invariance of the {Lagrangian} are a result of the field equations. When this is done the connection forms and the vierbein appear as a single g-bein on $\rm\tilde G$ (g being the dimensions of G) which also play the role of connections in computing covariant derivatives and curvatures.\\ On the other hand, implementing  the equations of motion derived from the action principle  of (\ref{action1}) one finds in a dynamical way that the theory lives effectively on a fiber bundle {$[\rm\tilde G/\rm H, \rm H]$}.

Note that even if in pure gravity the Hodge duality operator does not appear, in more general theories, like matter coupled gravity, it is precisely the absence of the Hodge duality operator which implies  that the choice of $\mathcal M_4$ turns out to be irrelevant, {Actually} any other $\mathcal M^\prime_4$ could work equally well and the physics would be the same on any of them.
Indeed a diffeomorphism in the direction orthogonal to space-time, considered from a \emph{passive} point of view, can be considered as a lifting of the hypersurfaces $\mathcal M_4$ to another  hypersurfaces $\mathcal M_4^\prime$ and  corresponds to the same theory in a different Lorentz frame. In other words  the lifting from $\mathcal M_4$ to  $\mathcal M^\prime_4$  corresponds to a Lorentz transformation. From an {\emph{active}} point of view, however, restricting the Lorentz factorized theory to a fixed $\mathcal M_4$, identified as the physical space-time,  the same diffeomorphism  corresponds to consider the theory together with all possible Lorentz transformations. Indeed{,} as we will explain in the next subsection{,} a diffeomorphism on the group manifold reduces to a Lorentz gauge transformation when the curvatures along the Lorentz subgroup are horizontal.

 Let us note that the requirements of geometricity given before for treating the simple case of pure Poincar\'e gravity {work} equally well in more complicated theories like matter coupled gravity theories in four or even higher dimensions.

In all these more realistic cases  the  requirement of the absence of the Hodge duality operator seems, at first sight, to be too strong.
 Indeed when any extended gravity theory  is coupled to scalars or vector fields their kinetic terms {require} the presence of the Hodge duality operator. For example, a kinetic term for a vector field should be written as proportional to
\begin{equation}\label{vector}
  {-} \int F_{\mu\nu}\,F^{\mu\nu} \sqrt{-g}d^4x= \frac12\int F\wedge * F {,}
 \end{equation}
 where $F_{\mu\nu}= \partial_{[\mu}A_{\nu]}$, $F=F_{ab}\,V^a\,V^b$ and $*F=\frac12 F^{ab}\epsilon_{abcd}V^c\,V^d$.
This apparent obstacle can be easily overcome introducing a \emph{first order formalism} for the vector field. Namely, we introduce  a 0-form {antisymmetric} Lorentz tensor $\hat F_{ab}$ {($\hat F_{ab}=-\hat F_{ba}$)} and write
for the kinetic term of the 2-form F the following {action:}
\begin{equation}\label{firstorder}
\mathcal{A_{\rm vec}} =- \int \hat F^{ab}\,\hat F_{ab}\Omega +\alpha \int {\hat F^{ab}\,F\,V^c\,V^d} \epsilon_{abcd},
\end{equation}
where $\Omega$ is the four{-}dimensional volume element $(-\frac{1}{4!}\epsilon_{pqrs}V^p\,V^q\,V^r\,V^s)$.
Varying the 0-form $\hat F_{ab}$ we find that choosing $\alpha =-\frac{1}{2}$ {we obtain}
\begin{equation}
\hat F_{ab}=F_{ab} ,
\end{equation}
where $F_{ab}$ are the components along two vielbein of the 2-form $F$.
Varying next the gauge field $A$,  from the second term of the action (\ref{firstorder}) we find the usual equation of motion:
\begin{equation}\label{secondorder}
\mathcal{D}_a F^{ab}=0 \rightarrow \mathcal{D}_\mu F^{\mu\nu}=0 .
\end{equation}
  In this way we see that using a first order formalism for the vector fields {Lagrangian}, the kinetic term can be obtained from the equation of motion starting from a geometric {Lagrangian} where the Hodge operator is absent. The same trick can be obviously used for the kinetic term of scalar fields.

 Quite generally the same requirement of geometricity should be made for any extension of the theory with additional fields (matter coupled gravity) and in particular in supergravity where, besides the mechanism of Lorentz coordinates factorization, the requirement of a \emph{geometric {Lagrangian}} turns out to be  necessary in order to implement in a geometrical way local supersymmetry. It is precisely this fact that makes the rather academic result of obtaining the factorization of Lorentz coordinates as a result of the field equations a simple standard example  to understand in a similar way the supersymmetry transformations in supergravity.

\subsection{ Gauge Transformations and Diffeomorphisms.}

It is interesting at this point to show in an explicit way how a diffeomorphism  reduces to a gauge transformation when the curvatures are horizontal, while it differs by curvature terms in the general case. We perform the derivation in a general group{-}theoretical setting so that it may apply to any (softened ) group or supergroup $\rm\tilde G$.\\
An infinitesimal element of the  GCTG   on $\rm\tilde G$ is given by a tangent vector on  $\rm\tilde G$, $\vec t=\epsilon^M T_M$, with $ \epsilon^M=\delta x^M {,} $
where the middle alphabet {L}atin capital indices are coordinate indices on  $\rm\tilde G$.
Using the vielbein $\mu^A$ of the whole (soft) group  $\rm\tilde G$ we can rewrite a tangent vector $\vec t$ as follows:
\begin{equation}\label{rewrite}
 \epsilon = \epsilon^{A} \tilde{T}_A,
\end{equation}
where $\epsilon^A= \epsilon^M \mu^A_M$  , and  $\tilde{T}_A= {T}_M \,\mu^M_A$. Here $\tilde{T}_A$ is the vector field generator dual to the non left-invariant 1-form $\mu^A$, $\m^A(\tilde T_B)=\delta ^A_B, $ and $\epsilon^A = \delta x^A$ is the infinitesimal parameter associated to the shift.
An  infinitesimal generator of  diffeomorphisms generated by $\epsilon^A$ is given by the Lie derivative
\begin{equation}\label{lie}
{\ell}_{\epsilon} \mu^A= \left(\iota_{{\epsilon}} d + d \iota_{{\epsilon}}\right) \mu^A,
\end{equation}
where $\iota_{{\epsilon}}$ is the contraction operator along $\epsilon$.\\
On the other hand the Lie derivative (\ref{lie}) can be also  rewritten as follows:
\begin{eqnarray}
{\ell}_{\epsilon} \mu^A &=& \left(\iota_\epsilon d + d \iota_\epsilon \right) \mu^A = \nonumber \\
&=& \iota_\epsilon d \mu^A + d \left( \iota_{\left(\epsilon^B\tilde{T}_B\right)} \mu^A \right) = \nonumber \\
&=& \iota_\epsilon d \mu^A + d \epsilon^A \ .
\end{eqnarray}
Adding and subtracting $C^A_{\phantom{A}BC} \mu^B \wedge \mu^C$ to $d \mu^A$ and using the definition of the covariant derivative
\begin{equation}\label{coventry}
  \nabla \epsilon^A = d\epsilon^A +C^A_{\phantom{A}{BC}}\mu^B\,\epsilon^C,
\end{equation}
 we find :
\begin{equation}\label{Liederiv}
{\ell} _{\epsilon} \mu^A = \iota_\epsilon \left(d \mu^A + \frac{1}{2}C^A_{\phantom{A}BC} \mu^B \wedge \mu^C\right) - \epsilon^B C^A_{\phantom{A}{BC}} \mu^C + d \epsilon^A.
\end{equation}
where we have used the antisymmetry of $C^A_{\phantom{A}{BC}}$ in the lower indices.
The terms in brackets define the curvature $R^A$ while the other two terms, using the antisymmetry of the structure constants in $(B,C)$ define the gauge covariant differential of $\epsilon^A$.
Therefore, using the \emph{anholonomized} parameter\footnote{By anholonomized parameter we mean that we are using the rigid group index of the vielbein $\mu^A $.} $\epsilon^A${,} the Lie derivative can be written as follows:
\begin{equation}\label{lie2}
{\ell}_{\epsilon} \mu^A= \left(\nabla \epsilon\right)^A + \iota_\epsilon R^A \,.
\end{equation}
Hence \textit{an infinitesimal diffeomorphism on the  manifold  $\rm\tilde G$ is a { $\rm G$}-gauge transformation plus curvature correction terms}.

In particular{,} if the curvature $R^A$ has vanishing projection along a  vector $\epsilon^{\tilde B}T_{\tilde B}$, where $\tilde B$ is an adjoint index of the subgroup $\rm H\subset \rm\tilde  G$ so that
\begin{equation}
\iota_\epsilon R^A \equiv \epsilon^{\tilde B} R^A_{\phantom{A}{\tilde B}C} \mu^C = 0,
 \end{equation}
then \emph{the action of the Lie derivative ${\ell}_{\epsilon}$ coincides with a gauge transformation}. In this case we recover the result that the  curvatures are \emph{horizontal} along the H directions and  the group manifold acquires the structure of a principal fiber bundle whose base manifold is $\rm\tilde G/ \rm H$, and $\rm H$ the gauge group. \\
 In conclusion{,} if the theory on  $\rm\tilde G$ can predict horizontal curvatures, it lives on  $[\rm\tilde G/H,\rm H]$ and it is equivalent to the original Cartan approach. This is in fact what we have found in the case of the Poincar\'e group. \\
We stress once again that the derivation of the formula AGCT, equation (\ref{lie2}), makes no explicit reference to the specific group  $\rm\tilde G$. It holds for any group, including supergroups, as we shall see in the supergravity case.

\section{ Geometric Supergravity.}

The fact that the fiber bundle structure in H of a gravity  theory can be obtained dynamically from a suitable action principle
 is certainly an interesting feature of these theories, since it sheds light on the geometrical origin of the theory and on the power of the action principle. However, from a purely physical point of view it does not seem to add anything important to our understanding of the theory. After all to write a theory possessing \emph{ab initio} a fiber bundle structure does not change anything in the development of the theory and in its physical results.\\
  The value of the previous detailed description of factorization of Lorentz parameters lies in the possibility to give a geometrical interpretation of supersymmetry analogous to the geometrical mechanism of factorization of the Lorentz coordinates  shown in the pure gravity case.\\ Indeed we will show, using  the simple example of ${\mathcal{N}}=1$, $D=4$ supergravity, that the invariance of the supergravity {Lagrangian} is due to the behavior of the supergroup curvatures along the fermionic components of the  supervielbein in superspace. Indeed, while the curvatures in the direction of the {``}Lorentzian" vielbein $\omega^{ab}$ of the supergroup have to be zero in order to obtain a fiber bundle structure for the H subgroup, exactly as it happens in gravity, this will not happen for the supercurvatures in the direction of the fermionic vielbein $\psi^\alpha$.\footnote{Here and in the following $\alpha$ is a spinor index in the relevant representation of SO(1,D-1).}\\
  What actually happens is the following: The dependence of the fields on the supergroup or superspace odd coordinates $\theta^\alpha$, does not imply their complete factorization, rather one finds that such components of the {curvature} 2-forms   can be expressed algebraically, actually \emph{linearly}, in terms of the curvatures restricted to the bosonic cotangent plane of  the embedded space-time hypersurface, namely in terms of $V^a\,V^b$, the basis on the cotangent space of ordinary space-time. Inserting this result into the Lie derivative formula (\ref{lie2}), one obtains a geometrical interpretation of the local supersymmetry transformations.\\
  For the sake of  brevity and simplicity we will show how this happens in  the simple example of pure ${\mathcal{N}}=1$, $D=4$ supergravity . However, the relevant results  hold exactly in the same way  for any supergravity theory, pure or matter coupled, in any dimension $ 4\leq D\leq 11$ and for any number  {$1\leq \mathcal{N} \leq 8$} of supersymmetry generators in the Lie {superalgebra}.

\subsection{${\mathcal{N}}=1$, $D=4$ Supergravity in the  Geometric Approach.}
We will now give the explicit description of the group manifold formalism for the ${\mathcal{N}}=1$, $D=4$ pure supergravity theory.

The graded group of ${\mathcal{N}}=1$, $D=4$ supergravity is the super-Poincar\'e group $\rm\tilde G=\overline{\rm O{S}p(1|4)}$, where the bar over $\rm O{S}p(1|4)$ means {In\"{o}n\"{u}}-Wigner contraction of the super Anti-de Sitter group (recall that $\rm SO(2,3)\simeq \rm Sp(4))$. The coadjoint multiplet of left-invariant 1-forms is
\begin{equation}\label{superad}
\sigma^A= (\mathring{\omega}^{ab},\mathring{V}^a,\mathring{\psi}^\alpha),
 \end{equation}
where $\mathring{\psi}^\alpha$ is  a Majorana spinor 1-form in the fermionic direction of the supergroup. The upper little ring means that, being left-invariant, they satisfy the Cartan-maurer equations ( vanishing curvatures):
\begin{equation}
d\sigma^A+\frac{1}{2}C^A_{\phantom{A}BC}\,\sigma^B\wedge \sigma^C=0.
\end{equation}
 These 1-form fields
 will now depend on the coordinates $\left(x^\mu, \eta ^{\mu\nu}, \theta^\alpha\right) $ of the supergroup. Deforming the $\sigma^A$ into the non left-invariant
  1-form $\mu^A=(\omega^{ab}{,}\,V^a\,,\psi^\alpha)$ allows us to define the following multiplet of super-curvatures{:}\footnote{Here and in the following we will mostly omit the spinor index $\alpha$ on the spinors. Moreover, if there is no risk of confusion, we shall often refer to the super-curvatures simply as curvatures.}
\begin{eqnarray}
 \label{defcur1} R^a_{\;\;b} &{\equiv}& d\omega^a_{\,\,\,b}- \omega^a_{\,\,\,c} \wedge\omega^c_{\,\,\,b} {,} \\
  \label{defcur2}  \hat T^a &{\equiv}& dV^a-\omega^a_{\,\,\,b} \wedge V^b -\frac{{\ii}}{2} \overline \psi \gamma^a \psi {,} \\
  \label{defcur3}    \rho &{\equiv}& \mathcal D \psi {= d \psi - \frac{1}{4} \gamma_{ab} \psi \omega^{ab}} ,
   \end{eqnarray}
   where we adopt a matrix  notation for the spinor current in equation (\ref{defcur2}) and we have denoted by $\rho$ the curvature 2-form  of $\psi${; $\gamma_a$ and $\gamma_{ab}$ are Dirac gamma matrices in four dimensions.} Moreover we have set a hat on the supertorsion, $\hat T^a$, to avoid confusion with the purely bosonic torsion $T^a$. The curvatures satisfy the following Bianchi identities:
\begin{eqnarray}
     \mathcal D R^{ab} &=& 0 {,} \\
     \mathcal D \hat T^{a}+ {R^{a}_{\;b}}\,V^b- {\ii} \bar \psi\gamma^a\, \rho &=& 0 {,} \\
    \label{susybianchi}\mathcal D \rho -\frac14 {\gamma_{ab}} \psi\, R^{ab}&=& 0 {.}
\end{eqnarray}
   Note that all  terms in the definition of the curvatures and in the Bianchi identities scale homogenously since $\omega^{ab},V^a, \psi$ and their curvatures have length scaling $[L^0],[L^1]$ and $[L^{1/2}]$, respectively.

In order to write down the {Lagrangian}  we require that it {is} \emph{geometric}. For the sake of clarity let us repeat here what it amounts to{,} adding some more obvious requirements:
\begin{itemize}
\item
It must be constructed using only differential forms, wedge product{s among them,} and the $d$ exterior differential;
\item
It must not contain the Hodge duality operator;\\
\item
As the Einstein term, which must be always present, {scales} as $[L^2]$, ({$[L^{D-2}]$ in $D$ dimensions}), all the terms must scale in the same way;
\end{itemize}
{To these requirements one usually adds the following one:}
\begin{itemize}
\item
If all the curvatures $R^A$ are zero (left-invariant $\sigma^A$ or vacuum configuration), the {Lagrangian} and the equations of motion must vanish identically.
\end{itemize}
This last requirement is
 useful in constructing {Lagrangian}s more complicated compared to the present {$\mathcal{N}=1$, $D=4$} theory.\\
It is easily seen that the only term we can add satisfying the first three requirements, for the $G=\overline{\rm O{S}p(1|4)}$ group, is the Rarita-Schwinger kinetic term (written in terms of differential forms). Thus we obtain
\begin{equation}\label{lagrsuper}
 \mathcal A_{D=4}^{{\mathcal{N}}=1}=  \frac{1}{4\kappa^2}\int_{\mathcal M_4 \subset {\overline {\textbf{OSp}(1|4)}}}\left[R^{ab}\,V^c\,V^d \epsilon_{abcd}+\alpha \overline \psi \gamma^5\gamma_a \mathcal D\psi\, V^a\right]\,.
\end{equation}
 where instead of a tilde we have used  a  boldface character to denote the soft supergroup manifold $\overline{\rm OSp(1|4)} $. This supermanifold has ten bosonic and four fermionic coordinates, namely $(x^\mu,\eta^{\mu\nu},\theta^\alpha )$.
The coefficient $\alpha$ between the Einstein and Rarita{-}Schwinger is related to the normalization of the gravitino {1-form} $\psi$ and will be fixed in a moment.
The equations of motion {obtained by varying} $\omega^{ab}{,} \, V^a{,}$ and $\psi$ are respectively:
\begin{eqnarray}
 \label{eqs1}  \epsilon_{abcd}\left(\mathcal D V^a +\frac{{\ii} \alpha}{8}\,\bar \psi\,\gamma^a\,\psi\right) \wedge V^d&=&0 {,} \\
\label{eqs2} 2 R^{ab}\wedge V^c\epsilon_{abcd}-\alpha \overline \psi \wedge\gamma^5\gamma_d\mathcal D\psi &=&0 {,} \\
 \label{eqs3} 2\gamma^5 \gamma_a \mathcal D\psi \wedge V^a -\gamma^5  \gamma_a\psi \wedge {\hat{T}}^a&=& 0\,.
\end{eqnarray}
  As the equations of motion have to vanish identically when all the (super{-})curvatures are zero (fourth requirement), we see that we must set in the left hand side of Eq.(\ref{eqs1}) $\alpha =4$ in order to have the super-torsion 2-form $\hat T^a$ as defined in (\ref{defcur2}). With this value of $\alpha$ equation (\ref{eqs1}) takes the form
\begin{equation}\label{tortor}
   \hat T^c \wedge V^d \epsilon_{abcd} = 0
\end{equation}
 and we see that when all the supercurvatures are zero the equations of motion vanish identically.\\

To analyze the content of equations (\ref{eqs2}), (\ref{eqs3}) and (\ref{tortor}) we expand the curvatures 2-forms  along the basis $\mu^A \wedge \mu^B$. Let us first consider their components along the basis 2-form containing at least on $\omega^{ab}$, namely (omitting {the} wedge symbol):
$$
\omega^{ab}\,\omega^{cd};\quad \omega^{ab}\,V^c;\quad \omega^{ab}\,\psi\,.
$$
It is then immediate to see that the equations of motion give \emph{horizontality } of all curvatures {$R^{ab}, \, \hat T^a, \, \rho$} in the Lorentz directions, namely
\begin{equation}\label{namely}
 R^A_{(ab)|(cd)}=R^A_{(ab)|c}=R^A_{(ab)|\alpha}=0 {,}
\end{equation}
exactly as in {the} pure gravity case. In this case, however, the supermanifold where the theory lives has a base space the ${\overline {\textbf{OSp}(1|4)}}/\rm SO(1,3)$ that is the \emph{softened super-coset of the vacuum $\overline{\rm OSp(1|4)}/SO(1,3) $}. We identify ${\overline {\textbf{OSp}(1|4)}}/\rm SO(1,3)$ as the \emph{superspace} ad will be denoted $\rm R^{(4|4)}$. Superspace has now as coordinates $x^\mu,\theta^\alpha$, since the Lorentz coordinates
$\eta^{\mu\nu}$ have been factorized.\\
However as the physical space-time $\mathcal M_4$ is four-dimensional while superspace
 has four extra fermionic dimensions, we are left with the same problem as in the bosonic case of pure gravity, namely to construct
 a suitable action for a {Lagrangian} 4-form in the eight-dimensional superspace. It is then natural to resort to the same procedure
 used for the pure gravity theory. Namely we can identify the space-time with  \emph{any} four-dimensional bosonic hypersurface $\mathcal M_4$
 embedded in superspace. Indeed, as the {Lagrangian} is completely geometrical and it does not contain the Hodge duality operator, the invariance under diffeomorphisms of the {Lagrangian} allows arbitrary deformations of $\mathcal M_4$ in superspace. Therefore the equations of motion, being independent of the particular hypersurface chosen, will hold on the full superspace. This is completely analogous  to the pure gravity case, the only difference being that the diffeomorphisms  on the bosonic group manifold have been replaced by diffeomorphisms in superspace.\footnote{If the factorization of the Lorentz coordinates have not yet been implemented then we  deform the hypersurface $\mathcal M_4$ on the full graded group manifold as it was done in the pure gravity case.}

 Since we  reduced ourselves to the study of a Lorentz invariant theory on superspace, {$\rm R^{(4|4)}$}, let us  shortly describe the  geometric structure of the theory  in superspace .

Since the Lorentz coordinates have already {been} factorized, on  {$\rm R^{(4|4)}$} the general base of 1-forms is given by the set of  super-vielbein $E^A$, namely $E^A= \left(V^a,\psi^\alpha\right)$, where $\psi^\alpha$, $\alpha =1,\dots 4$, is the fermionic vielbein, that {is a} Majorana spinor 1-form  named gravitino. The action (\ref{lagrsuper}) is now reduced to the following form:
\begin{equation}\label{actionsuper}
  \mathcal A= \frac{1}{4\kappa^2}\int_{\mathcal M_4 \subset \rm R^{(4|4)}}\left[R^{ab}\,V^c\,V^d \epsilon_{abcd}+ 4 \overline \psi \gamma^5\gamma_a \mathcal D\psi\, V^a\right]\,.
\end{equation}
Let us introduce for the sake of brevity the following notation. We denote by  $ R^A_{(p,q)}$, {$A=(ab,a,\alpha)$}, the components of the curvature along $p$ bosonic vielbein $V^a$ and $q$ fermionic vielbein $\psi$. Moreover we call \emph{outer} all the components were $q\neq 0$, that is those components having at last one index along the $\psi$ direction, while when $q=0$, that is when the only non vanishing components are along the bosonic vielbein,  they will be called \emph{inner.}

 To analyze the equations {(\ref{eqs2}), (\ref{eqs3}), and (\ref{tortor})}, now restricted to superspace,  we must expand the curvatures along a complete basis of 2-forms in superspace. In this case we have
\begin{equation}\label{partialb}
   E^A\wedge E^{{B}} = \{(V^a\,V^b);\quad (V^a\,\psi^\alpha); \quad (\psi^\alpha, \psi^\beta)\}.
\end{equation}
  Let us first work out equation (\ref{tortor}). Expanding $\hat T^a$ we have
\begin{equation}\label{exptor}
    \hat T^a= {\hat{\tilde{T}}^a_{bc}} {V^b \, V^c} + {\bar H^a_{\;c}} \psi\,V^c + \bar\psi K^a \,\psi {,}
\end{equation}
  where $\hat T^a_{(1,1)} = {H^a_{\;c}}$ is a spinor {and} $K^a$, defining the $\hat T^a_{(0,2)}$ component, is proportional to a {gamma matrix in four dimensions}. Considering equation (\ref{tortor}){,} one  easily conclude that the components of $\hat T^a_{(1,1)}$ must be zero, while $\hat T^a_{(0,2)}$ would only  change the normalization of the gravitino {1-form} in the definition (\ref{defcur2}), so we can put them also to zero. {Then, equation (\ref{tortor})} restricted to the  {$V\,V$ ($\hat T^a_{(2,0)}$)}  components  has exactly the same form as in the pure gravity case, {that is} (\ref{tor}), provided we replace $T^a$ with $\hat T^a$. Note, however, that in this case solving for the spin connection $\omega^{ab_\mu}$ with the usual procedure gives a spin connection which depends not only from the vielbein and their derivatives, but also from gravitino bilinears.\footnote{See e.g. reference  \cite{Castellani:1991et}.}
With exactly the same computations as those made for pure gravity, one easily {obtains} the  vanishing of the  {$\hat{\tilde{T}}^a_{bc}$}  components and therefore the {whole super-torsion 2-form} is zero.\\
In the same way{, in order} to solve the equations (\ref{eqs2}) and (\ref{eqs3}) we  expand the curvatures  $\rho$ and $R^{ab} $ along a complete basis of 2-forms in superspace.
  As $\hat T^a=0$ the equation (\ref{eqs3})  takes the form
\begin{equation}\label{psivrho}
   \gamma_{a}\mathcal D \psi\,V^a=0
\end{equation}
  and expanding $\rho\equiv \mathcal D\,\psi$ as
\begin{equation}\label{exprho}
    \rho^\alpha =\tilde\rho^\alpha_{ab}V^a\,V^b+ H_a \psi^\alpha\,V^a + \Omega_{\alpha\beta}\psi^\alpha\,\psi^\beta ,
\end{equation}
where, for the sake of clarity, we have made the spinor index explicit.
From equation (\ref{psivrho})  one easily realizes that $H_a=\Omega_{\alpha\beta}=0$,  so that the 2-form $\rho$ has only components {$\rho_{(2,0)}$} on the cotangent space of $\mathcal M_4$, namely
\begin{equation}\label{gravinocur}
   \rho={\tilde \rho_{ab}} V^a\,V^b\,.
\end{equation}
{We warn the reader that since we are now in superspace the rigid indices cannot be traded with coordinate indices using the bosonic vierbein $V^a_\mu$. Indeed, the full set of supervielbein is now given by $E^A= \left(V^a,\psi^\alpha\right)$ and we should invert the matrix $E^A_\mu$ to find the space time components.  This is in fact the reason why we have denoted with a tilde the components of the supercurvatures along two bosonic vierbein. A simpler way to find the space components is to project the equations on the space -time basis $dx^\mu\wedge dx^\nu$. For example from Eq. (\ref{exprho}), projecting on the space-time basis we obtain
 \begin{equation}\label{proj}
   \rho^\alpha_{\mu\nu} =\tilde\rho^\alpha_{ab}V^a_\mu\,V^b_\nu+ H_a \psi^\alpha_\mu\,V^a_\nu + \Omega_{\alpha\beta}\psi^\alpha_\mu\,\psi^\beta_\nu. 
 \end{equation}
 where the indices $\mu\nu$ are undestood to be antsymmetric. We see that the tilded components of $\tilde\rho_{\mu\nu}$ differ from the the real space time components $\rho_{\mu\nu}$ by terms in the gravitino fields, namely \emph{outer terms}. They are commonly named in the literature as \emph{supercovariant field strengths}.\footnote{The name \emph{supercovariant} means that their supersymmetry transformation law does not contain derivatives of the supersymmetry parameter $\epsilon^\alpha$.}
 However, in our case, as far the $\hat T^a$ and $\rho$ curvatures are concerned, we can easily convert rigid Lorentz indices in world indices as usual, since in the present case they do not have \emph{outer} components $(V {\wedge} \psi)$ and $ (\psi {\wedge} \psi)$. Then, for the components of the aforementioned curvature 2-forms we can neglect the tilde symbol. Instead, as we will see in a while and further discuss in the sequel, the space-time components of the Lorentz curvature do not coincide with the components along $(V^c\wedge V^d)$ expanded along the differentials of the coordinates.}
 \\
Indeed{,} expanding  $R^{ab}$,
\begin{equation}\label{decrab}
   R^{ab}= \tilde R^{ab}_{cd}V^c\,V^d +  \overline\Theta^{ab}_c\psi\,V^c+ \bar\psi K^{ab}\psi,
\end{equation}
from equation (\ref{eqs2}) we find
\begin{equation}\label{tetta}
{\overline\Theta^{ab}_c =- \epsilon^{abrs} \bar{ \rho}_{rs}\gamma^5\gamma_c - \delta^{[a}_c\epsilon^{b]mst}\bar{ \rho}_{st}\gamma^5 \gamma_m}
\end{equation}
and $K^{ab}=0$.

  Note that the value for the outer component of $R^{ab}_{(1,1)}=\overline\Theta^{ab}_c\psi\,V^c$ given in (\ref{decrab}) is written in terms of the components of the gravitino curvature along  $V^a\,V^b$, namely $ \rho_{ab}$. Thus an \emph{outer} component of the Lorentz curvature 2-form is linearly expressed, on-shell, in terms of the \emph{inner} components of the fermionic curvature $\rho$. This property is called \emph{rheonomy} and will be discussed more generally in the following. Physically it just means that no new {degree} of freedom {is} introduced in the theory other than those already present on space-time.  Actually, if rheonomy is assumed \emph{a priori}, and  we  take advantage of the fact that all the coefficients in the expansion along the supervielbein  must give rise to terms with same scale as the corresponding curvatures, then one easily {recognizes} that the previous results for the outer components of the curvature multiplet are easily recovered {(as for the possibility to have dimensional constant, see footnote 8)}.
   In conclusion{,} the solution of the equations of {motion} (\ref{eqs1}), (\ref{eqs2}), (\ref{eqs3}) for the \emph{outer} and \emph{inner} projections of the curvature multiplet gives:
\begin{eqnarray}
 \label{parcur}  R^{ab}&=& \tilde R^{ab}_{cd}V^c\,V^d + \overline\Theta^{ab}_c\psi\,V^c {,} \\
  \label{partor}  {\hat{T}^a} &=& 0 {,} \\
   \label{parrho} \rho &=& {\rho_{ab}}  V^a\,V^b\,.
\end{eqnarray}
 Finally, inserting  the parameterizations  (\ref{parcur}), (\ref{partor}), (\ref{parrho}) in the equations of motion (\ref{eqs1}), (\ref{eqs2}), (\ref{eqs3}),  we find that the components  of the equations of motion along {$V^a \, V^b\,V^c$, that is $ R^A_{(3,0)}$,} give the space-time equations of motion:
\begin{eqnarray}
   \tilde R^{am}_{bm} -\frac12 \delta^a_b  \tilde R^{mn}_{mn}&=& 0 {,} \\
  \hat T^a_{mn} &=& 0 {,} \\
 \label{spacetimeeq} \epsilon^{mnrs}\gamma^5\gamma_s  {\rho_{mn}} &=& 0 {.}
\end{eqnarray}

 To find the relation between the  components on space-time of these equations, as observed before instead of inverting the supermatrix $E^A_\alpha$ we project the equations  (\ref{parcur}), (\ref{partor}), (\ref{parrho}) on the space-time 2-form differentials $dx^\mu\wedge dx^\nu$. We obtain
\begin{eqnarray}
\label{supercov}  R^{ab}_{\mu\nu}&=& \tilde R^{ab}_{cd}V^c_\mu\,V^d_\nu + \overline\Theta^{ab}_c\psi_{[\mu}\,V^c_{\nu]} {,} \\
  \hat T^a_{\mu\nu} &=& \hat T^a_{bc} V^b_\mu\,V^c_\nu {,} \\
  \rho_{\mu\nu} &=& \rho_{ab}V^a_\mu\,V^b_\nu {.}
\end{eqnarray}
We have already seen that the space-time components of $\hat T^a$ and $\rho$ curvatures are obtained as usual converting rigid indices in curve ones using the bosonic vierbein $V^a_\mu$ since their parametrization does not contain $\psi $ fields. Instead the space{-time} components of the Lorentz curvature  expanded along the differentials of the coordinates, namely $R^{ab}_{\mu\nu}$ do not coincide with the components along $(V^c\wedge V^d)$. In fact writing equation (\ref{supercov}) as
\begin{equation}\label{supercov1}
 \tilde R^{ab}_{\mu\nu} = R^{ab}_{\mu\nu}- \overline\Theta^{ab}_c\psi_{[\mu}\,V^c_{\nu]} {,}
\end{equation}
where $\tilde R^{ab}_{\mu\nu}\equiv  R^{ab}_{cd}V^c_\mu\,V^d_\nu${,} we see that the Einstein equation of motion, written in terms of the $\tilde R^{ab}_{\mu\nu}${,} contains extra terms linear in the inner components {$\rho_{rs}{\equiv} \rho_{\mu\nu}\,V^\mu_r\,V^\nu_s$}. These {terms} give rise to the energy-momentum tensor of the gravitino field $\psi_\mu$.

\subsection{Invariance of the Lagrangian.}

Let us now check the supersymmetry invariance of the {Lagrangian}. {In the geometric approach, the supersymmetry invariance of the Lagrangian is expressed by the vanishing of the Lie derivative of the Lagrangian for infinitesimal diffeomprphisms in the fermionic directions of superspace.}
Using the Lie derivative with a tangent vector
 $\vec \epsilon= \epsilon^\alpha \vec D_\alpha${,} where $\vec D_\alpha$ is the tangent vector dual to $\psi^\beta$, we must have
\begin{equation}\label{bob}
  \delta_\epsilon \mathcal L\equiv \ell_\epsilon \mathcal L={\iota_{ \epsilon}}d\mathcal L= 0 ,
\end{equation}
where  we have discarded the total derivative term {$d (\iota_{\epsilon}\mathcal L)$} and other possible exact 4-forms on the right hand side as we are assuming that the fields  vanish at infinite so that any exact form does not contribute to the action.\footnote{Note that the left hand side of equation {(\ref{bob})} is not zero since we are now in the {(4+4)-}dimensional superspace.}
Taking into account the definition (\ref{defcur2}), a simple computation gives
\begin{eqnarray}\label{invar}
 &&d\mathcal L= 2  R^{ab}\,\hat T^c \,V^d\epsilon_{abcd} + {\ii}\,R^{ab}{\bar \psi}\,\gamma^c\psi \,V^d \epsilon_{abcd} +\nonumber\\
  &&+4 \bar \rho\,\gamma^5 {\gamma_a} \rho V^a + \bar \psi\gamma^5 {\gamma_c\,\gamma_{ab}} \psi\,R^{ab}\,V^c  -4 \bar\psi\gamma^5 {\gamma_a} \rho \,\hat T^a -2\, {\ii} \bar \psi \gamma^5 {\gamma_a} \,\rho\,\psi \gamma^a \psi,
\end{eqnarray}
 where we have used the Bianchi identities and $\iota_\epsilon \psi=\epsilon, \, \iota_\epsilon V^a=0$.
Using the Fierz identity for 1-form spinors $\gamma^a \psi\bar \psi\,\gamma_a\,\psi=0$, (see reference \cite{Castellani:1991et}) and performing the gamma matrix algebra{,} one finds:
\begin{equation}\label{dL}
  d\mathcal L=R^{ab}\,\hat T^c \,V^d\epsilon_{abcd}+ \bar \rho\gamma^5 {\gamma_a} \rho V^a -4\bar\psi\gamma^5 {\gamma_a} \rho {\hat T^a} \,.
\end{equation}
  Finally{,} contracting with the tangent vector $\epsilon^\alpha \vec D_\alpha${,} we obtain
\begin{eqnarray}\label{invar2}
 &&\iota_{\vec \epsilon}\,d\mathcal L= 2 (\iota_\epsilon R^{ab})\,\hat T^c \,V^d\epsilon_{abcd} + 2R^{ab}(\iota_\epsilon\, \hat T^c) \,V^d \epsilon_{abcd} {+}
   8(\iota_\epsilon \bar \rho)\gamma^5 {\gamma_a}\rho V^a \nonumber\\
  &&- 4\bar \epsilon\gamma^5 {\gamma_a} \rho\,\hat T^a -4 \bar \psi\gamma^5 {\gamma_a} (\iota_\epsilon\rho) \hat T^a {-}4 \bar\psi\gamma^5 {\gamma_a} \rho  (\iota_\epsilon \,\hat T^a)= d( \rm 3-form).
\end{eqnarray}

From {(\ref{invar2})} we see that we can have an invariant action if we require constraints on the components of the curvatures. Indeed, if we set
\begin{equation}\label{consrta1}
 \iota_\epsilon T^a=0;\quad \iota_\epsilon \rho =0
\end{equation}
and furthermore
\begin{equation}\label{consrta2}
 2\left(\iota_\epsilon R^{ab}\right)\,{V^d}\,\epsilon_{abcd}+4\bar \epsilon\gamma^5 {\gamma_c} \rho  =0\,.
\end{equation}
we find $\delta_\epsilon \mathcal L =0$, that is \emph{invariance of the Lagrangian under supersymmetry}.

We note that the requirements  {(\ref{consrta1})} and   {(\ref{consrta2})} are the same   of the on-shell constraints (\ref{partor}) and (\ref{parrho}) found from the equations of motion. In particular {(\ref{consrta2})} gives the  solution
\begin{equation}\label{ex}
  \iota_\epsilon R^{ab}=\bar \epsilon \Theta^{ab}_c V^c {,}
\end{equation}
where $ \Theta^{ab}_c$ has been defined in equation (\ref{tetta}).\\
In other words we retrieve exactly the same constraints on the curvatures as those found from the equations of motion.

 We conclude that the \emph{supergravity Lagrangian is invariant under (local) supersymmetry transformations when the superspace curvatures are defined by the equations (\ref{parcur})-(\ref{parrho})}.
However, this restricted form of the curvatures in superspace imply that the supersymmetry transformations given below leaving the Lagrangian invariant do not form a closed algebra, unless one uses the equations of motion.\footnote{As it is well known there exist theories in {$D=4$ and $D=5$} which admit \emph{auxiliary fields}, that is fields that added to the coadjoint supermultiplet make the supersymmetry transformations, besides leaving the Lagrangian invariant, to close the supersymmetry algebra \emph{off-shell} . This is related to the fact that their introduction pairs the number of off-shell degrees of freedom between boson and fermions. Moreover they are not dynamical as their equations of motion make them to vanish. However, it does not {seem} possible to extend their introduction to higher{-}dimensional theories nor to matter coupled supergravities. Therefore we do not treat them in this short review.} This is best understood looking at the supersymmetry transformation laws.

Indeed, since we have found the \emph{on-shell} value of the curvatures, we may apply the Lie derivative formula (\ref{lie2}) to write down the superspace diffeomorphisms of the gauge fields ${\omega^{ab},V^a,\psi}$ using a generic tangent vector $\vec \epsilon= \epsilon^{ab}D_{ab} + \epsilon^a D_{a} +\epsilon^\alpha D_\alpha$, where the tangent vectors $D_{ab}\,D_{{a}},D_\alpha$ are dual to the gauge field 1-forms {$\omega^{ab},\,V^a,\,\psi^{\alpha}$}. We find:
\begin{eqnarray}
 \delta_\epsilon \omega^{ab} &=&(\nabla \epsilon)^{ab} + \epsilon^r V^s R^{ab}_{rs}+   \overline\Theta^{ab}_r \psi \epsilon^r +   \overline\Theta^{ab}_r \epsilon V^r {,} \\
  \delta_\epsilon V^a &=& (\nabla \epsilon)^{a} {,} \\
  \delta_\epsilon \psi^\alpha  &=& (\nabla \epsilon)^\alpha+ \epsilon^r \rho_{rs}^\alpha V^s.
\end{eqnarray}
Restricting ourselves to the Lie derivative along the fermionic supersymmetry parameter $\epsilon$ only,  that is setting $\epsilon^{ab}=\epsilon^{a}=0$, we have
\begin{eqnarray}
\label{susyosp1} \delta_\epsilon \omega^{ab} &=&(\nabla \epsilon)^{ab} + \overline\Theta^{ab}_r \epsilon V^r {,} \\
\label{susyosp2} \delta_\epsilon V^a &=& (\nabla \epsilon)^{a} {,}  \\
 \label{susyosp3} \delta_\epsilon\psi^\alpha  &=&(\nabla \epsilon)^\alpha.
\end{eqnarray}
Here {the} symbol $\nabla$ denotes the \emph{gauge covariant derivative} of the coadjoint multiplet of $\overline{\rm O{S}p(1|4)}$. The Lorentz content of the gauge covariant derivative when acting on the $\overline{\rm O{S}p(1|4)}$ adjoint multiplet can be read off directly from the Bianchi identities (\ref{susybianchi}). Indeed both the parameters $\epsilon^A$ and the curvatures are in the coadjoint multiplet of the supergroup. Therefore:
\begin{eqnarray}
 \delta_\epsilon^{(gauge)} \omega^{ab} &=&  (\nabla \epsilon{)}^{ab}=\mathcal D {\epsilon^{ab}} {,} \\
 \delta_\epsilon^{(gauge)} V^a &=& (\nabla \epsilon)^{a}=\mathcal D \epsilon^{ab} + \epsilon^{ab} V_b -{\ii}\bar \psi \gamma^a \epsilon {,} \\
  \delta_\epsilon^{(gauge)} \psi&=&(\nabla \epsilon)= \mathcal D \epsilon -\frac14 \epsilon^{ab}\gamma_{ab}\psi {,}
\end{eqnarray}
where $\mathcal D$ denotes the Lorentz covariant derivative.
Setting again  $\epsilon^{ab}=\epsilon^{a}=0$ and substituting in (\ref{susyosp1}),\,(\ref{susyosp2}),\,(\ref{susyosp3}) we find the final form of the supersymmetry transformations:
\begin{eqnarray}
 \label{finsusyR}\delta_\epsilon \omega^{ab} &{=}&   \overline\Theta^{ab}_r \epsilon  V^r\\
\label{finsusyT} \delta_\epsilon V^a &=&  -{\ii}\bar\psi \gamma^a \epsilon {,} \\
\label{finsusyro}\delta_\epsilon\psi  &=& {\mathcal D \epsilon\,} .
\end{eqnarray}

Now we recall that the Lie derivative along tangent vectors $\tilde T _A$ satisfy an algebra isomorphic to the Lie algebra of the vector fields
$[\tilde {T}_A,\tilde {T}_B]=\left(C^A_{\;BC}+R^A_{\;BC}\right)\,\tilde T_C$, namely
\begin{equation}\label{liederalgebra}
  [{\ell}_{\tilde T_A},{\ell}_{\tilde T_B}]=\ell{_{[\tilde T_A, \tilde T_B]}} ,
\end{equation}
if the supercurvatures $R^A_{\;BC}$ are completely general, that is if they do not satisfy any constraint. In our case they satisfy the constraints (\ref{parcur})-(\ref{parrho}) and in general the Lie derivative algebra, namely the algebra of supersymmetry transformations, \emph{cannot close off-shell}.\footnote{Unless the constraints coincide with  horizontality of the full set of curvatures as it happens for the Lorentz gauge invariance.} Actually, as will be discussed in subsection 3.3, requiring that the  Bianchi identities on the constrained curvatures be satisfied, one finds that their components on the bosonic cotangent plane $R^A_{rs}$ satisfy the equations of motion of the theory. It follows that the supersymmetry algebra of the transformations leaving the Lagrangian invariant, associated to the two tangent vectors $\epsilon^\alpha\,D_\alpha$,  will in general only \emph{close on-shell}, that is, only if the equations of motion are satisfied.

\subsection{ Supersymmetry as Diffeomorphisms in Superspace and Rheonomy.}

Let us now discuss the results obtained so far.
\begin{itemize}
\item
Even if the supercurvatures {$\hat{T}^a$} and $\rho$, equations (\ref{partor}) and (\ref{parrho}){, respectively,} have no components along the fermionic vielbein $\psi$, a non{-}vanishing component along $\psi {\wedge} V^a$ does appear in the on-shell value of the Lorentz supercurvature{, that is} (\ref{tetta}). \emph{This is sufficient to exclude factorization of the odd fermionic coordinates}. Indeed its presence makes the supersymmetry transformation a \emph{diffeomorphism} in superspace and \emph{not a gauge transformation}.\\
It must also be noted  that  the absence of such fermionic components in the (on-shell) gravitino curvature $\rho$ implies that the supersymmetry variation of $\psi$, given in equation (\ref{finsusyro}){,} makes the transformation of the gravitino gauge field the same as if the {Lagrangian} were  invariant under supersymmetry gauge transformations.  However, as the supersymmetry transformations of the Lagrangian do not close an algebra, the gravitino transformation law is actually a diffeomorphism, and the {Lagrangian} cannot be a true gauge symmetry  because of the absence of factorization, as we have  previously shown.\footnote{Note that if the {Lagrangian} were invariant under supersymmetry gauge transformations the superfields would only depend on the $x^\mu$ coordinates.}\\ The point is that  such behavior of the gravitino transformation law is due to  the very simple form of the minimal ${\mathcal{N}}=1$, $D=4$ pure supergravity. Any other supergravity with ${\mathcal{N}}>1$ or $D>4$ or even the same theory ${\mathcal{N}}=1$, $D=4$ coupled to matter multiplets exhibits a gravitino curvature with components ${\rho_{(1,1)}}\neq 0$ so that the $\delta_\epsilon \psi$ will have, besides the Lorentz covariant derivative of the supersymmetry parameter{,} also terms along {$\psi \wedge V^a$}.

 As an  example, let us consider ${\mathcal{N}}=2$, $D=4$  pure supergravity. Here the supergroup is $\overline{\rm O{S}p(2|4)}$. The coadjoint gauge supermultiplet is now given by $\mu^A=(\omega^{ab}, V^a,\psi_A, \mathcal A)$, where $\mathcal A$ is {a} $\rm U(1)$ gauge field 1-form  and the index A enumerates the  gravitinos in the two{-}dimensional representation of $\rm SO(2)$. The definition of the associated supercurvatures are obtained as always starting from the Maurer-Cartan equations dual to the algebra of the (anti{-})commutation generators {and} deforming the left-invariant 1-forms into {non} left-invariant ones. Without giving the derivation, we write, besides {the} definitions of the supercurvatures on the left hand side,  also their on-shell parametrization  as found from the analysis of the equations of motion:
\begin{eqnarray}
R^{ab} &{\equiv}& d\omega^a_{\,\,\,b}+ \omega^a_{\,\,\,c}\omega^c_{\,\,\,b} \nonumber \\
& =& \tilde R^{ab}_{cd}V^c V^d + \overline\Theta^{ab}_{A|c}{\psi^A} V^c -\bar\psi_A\left(F^{ab}+{\ii} ^* F^{ab} \gamma^5\right)\psi_B {\epsilon ^{AB}} {,} \\
{\hat T}^a &{\equiv}& \mathcal D V^a -\frac{{\ii}}{2} \bar\psi_A \gamma {\psi^A}=0 {,} \\
F &{\equiv}& \mathcal F +  {\epsilon ^{AB}} \bar\psi_A \psi_B =F_{ab} V^a V^b {,} \\
\rho_A &{\equiv}& \mathcal D\psi_A = \tilde\rho_{A|ab} V^a V^b +\left( \gamma^a \,F_{ab} + {\ii} \gamma^5 \gamma^a {^*} F_{ab}\right){\epsilon _{AB}\bar\psi^B} V^b {,}
\end{eqnarray}
where $\mathcal F=dA$ and $F$ is the supercurvature{,} and  ${^*} F_{ab}$ is the Hodge dual of $F_{ab}$. \\
  The important thing to note is that the parametrization of the {curvature} 2-forms given by the equations of {motion} are all given in terms of {their inner components}, namely $\tilde R^{ab}_{cd}$, {$\tilde\rho_{A|ab}$}, and {$F_{ab}$} (${\hat T}^a_{bc}$ is zero).\footnote{Note that {$F_{ab}$} has no tilde since {$F$} has components only along $V^a \, V^b$.}

  Since the on{-}shell values of the supercurvatures is known, the supersymmetry transformation laws of the coadjoint supermultiplet, now containing also $\mathcal A$, can be obtained at once from the general formula (\ref{Liederiv}). Looking at the Lie derivative formula, we see that the transformation laws of the multiplet of fields can be simply obtained  performing the contraction of the on-shell curvatures with respect to the tangent vector $\bar{\epsilon}\,D$ and adding to the gravitino transformation the Lorentz covariant derivative of the supersymmetry parameter as it happens in the $\overline{\rm O{S}p(1|4)}$ case.  We find:
\begin{eqnarray}
 \label{finsusyR2} \delta_{\epsilon} \omega^{ab} &=& \overline\Theta^{ab}_{A|r} {\epsilon^A} V^r {,} \\
\label{finsusyT2} \delta_{\epsilon} V^a &=&   - {\ii\bar{\psi}_A \gamma^a \epsilon^A} {,} \\
\label{finsusyro2}\delta_\epsilon\psi_A  &=&\mathcal D \epsilon_A+ {\ii} \,\epsilon_{AB}F^{ab} V^b \gamma^a\epsilon_B+ {\ii} \,\frac12\epsilon_{AB}\epsilon_{abcd}F^{cd} V^b \gamma^5 \gamma^a {\epsilon^B} {,} \\
\label{finsusyF}\delta_\epsilon \mathcal A &=& 2 {\epsilon^{AB}} \bar\psi_A\,\epsilon_B {.}
\end{eqnarray}
From this example we see that in general not only the Lorentz curvature $R^{ab}$, but also the other supercurvatures have non-vanishing components along the $\psi$-directions.
\item
We can now resume our analysis in the following way: \\
 Supersymmetry can be interpreted geometrically as the requirement that the superspace equations of motion imply that \emph{the outer components of the super-curvatures are expressible algebraically (actually linearly) in terms of the components along two inner vielbein}. As already mentioned this property has been called \emph{rheonomy}. Note that \emph{rheonomy  is just a geometrical interpretation of supersymmetry originally introduced on space-time}.
  Explicitly, the occurrence of \emph{rhenomy} can be written as follows:
\begin{equation}\label{algebraic}
    R^A_{\alpha\,C}= C^{A|mn}_{\alpha\,C|B}\, R^B_{mn} {,}
\end{equation}
  where $C^{A|mn}_{\alpha\,C|B}$ are suitable invariant tensors of  the supergroup S$\rm\tilde G$ defining the basic superalgebra on which the theory is constructed, S$\rm\tilde G$=$\overline {\rm O{S}p(1|4)}$  in our case.  The geometric meaning of this property can be better understood if we use the Lie derivative formula (\ref{lie2}) in superspace. Inserting (\ref{algebraic}) in the Lie derivative formula (\ref{lie2}) for a  supergroup S$\rm\tilde G$ we obtain:
\begin{equation}\label{rhe}
    \delta \mu^A = (\nabla \epsilon)^A + 2{\bar{\epsilon}} \,C^{A|mn}_{\alpha\,C|B}\, R^B_{mn}.
\end{equation}
 On the other hand{,} the Lie derivative can be interpreted either from the \emph{passive} or  from {the} \emph{active} point of view. From the passive point of view the supersymmetry transformation
 along the $\epsilon^\alpha=\delta \theta^\alpha $ parameter is interpreted as the lift from a given $\mathcal M_4$ to an infinitesimally close $\mathcal M^\prime_4$ which does not change the physical content of the theory, since it is described by the same Lagrangian after a supersymmetry transformation (and  a Lorentz gauge transformation) has been made.\footnote{The passive interpretation of the Lie derivative explains the world \emph{rheonomy} given to this geometrical interpretation of supersymmetry. Indeed, referring to the lift $\mathcal M_4 \rightarrow \mathcal M^\prime_4$, in ancient $\rm\tilde G$reek {``}rhein" means flow and {``}nomos" means law{.}} From the active point of view, however, it transforms a given configuration on $\mathcal M_4$, which we can take as space-time, setting $\theta^\alpha=\delta\theta^\alpha=0$, to another physically equivalent configuration  on the same space-time hypersurface.
 This property allows us  to restrict the theory, {the Lagrangian, and the equations of motion} to any such arbitrarily chosen  hypersurface $\mathcal M_4 \, (\theta^\alpha=d\theta^\alpha=0)$ embedded in superspace and identified with space-time.
\end{itemize}

One can  now appreciate why we have illustrated in detail the mechanism of the Lorentz coordinate factorization in the gravity case defined on the Poincar\'e manifold.\\ Actually the interpretation of the rheonomy mechanism just illustrated  is quite analogous to the interpretation of Lorentz transformations for gravity constructed directly on a group manifold. Indeed, in the case of pure gravity,  we have seen  that a transfer of information from any $\mathcal M_4\subset {\rm\tilde G}$ to any other $\mathcal M^\prime_4\subset {\rm\tilde G}$ implies a $\rm SO(1,3)$ transformation or{,} equivalently, a change of Lorentz configuration on the fixed space-time hypersurface.\\
 On the other hand,in our example of ${\mathcal{N}}=1$, $D=4$ supergravity, 
besides  deducing the factorization of the Lorentz coordinates  exactly as  in the pure gravity case, we have further illustrated that the equations of motion allow us to deduce that the  transfer of information concerns not only Lorentz gauge transformations but, what is our main goal, also \emph{supersymmetry}.\\
However the difference between $\rm SO(1,3)$ transformations and supersymmetry is that in the first case, due to the horizontality of the curvatures in the  Lorentz directions, $\omega^{ab}$ the supergroup $\rm\tilde G$ acquires the structure of the fiber bundle $[\rm\tilde G/H,H]$, and the Lie derivative reduces to a Lorentz gauge transformation. On the other hand, in the case of supersymmetry,  curvatures are not horizontal along the $\psi$  gauge fields and the Lie derivative gives the geometric interpretation of supersymmetry.  Actually, it is the rheonomic mechanism one is interested in, and in fact, quite generally,  in the construction of any supergravity theory the fiber bundle structure with a Lorentz fiber is assumed a priori as it can be considered of academic interest to obtain it from the variational principle.. In a way, restricting a supergravity theory  to a {factorized superspace $\rm\tilde G/\rm SO(1,3)$
 includes on a $\mathcal M_4$ slice, identified as space-time, all possible supersymmetry related Lagrangians.

In conclusion, the entire physics is contained in any single $\mathcal M_4$ or, equivalently, the supersymmetry transformations relate the fields on $\mathcal M_4$ to the fields on any other submanifold $\mathcal M^\prime_4$. It must be kept in mind that, since supersymmetry is a Lie derivative (super-diffeomorphism) in superspace, it \emph{is not a gauge symmetry}. Indeed, as we have seen, its algebra does not even close off-shell.

\subsection{{Building Rules of a General Lagrangian.}}

Our previous detailed examples give us simple rules for the construction of a general {(geometric)} {Lagrangian} in {$D$} dimensions:
\begin{itemize}
\item Starting from the Cartan-Maurer equations of a supergroup one defines the supercurvatures  of the supergroup
 on which the theory is based. More precisely one  writes the dual form of its super-Lie algebra and deforms the left-invariant 1-forms so that we can define the super-curvatures.
 \item
The {Lagrangian} must be a {$D$}-form built in terms of the 1-form gauge fields $\mu^A$ {and} their curvatures $R^A$. It must contain the Einstein term  $R^{ab} V^{c_1}\dots V^{c_{D-2}}\epsilon_{abc_1\dots c_{D-2}}$ whose scale is $[L^{D-2}]$. All the other terms should also scale as the Einstein term. At this point the Lagrangian contains a set of undetemined coefficients wich will be fixed from the superspace equations of motion.
\item
The Hodge operator must be absent; the kinetic terms of scalar and vector fields must be written in first order formalism, as shown in the example of Section 2.1.
\item
All the possible terms satisfying the previous requirements must be present.
\item
If gauge invariance under an H subgroup of $\rm\tilde G$ is imposed a priori,  where H  is the gauge group of the theory containing the Lorentz group as a factor $\rm H=\rm SO(1,D-1){\otimes} {\rm H}^\prime$, the action is obtained by integrating the {Lagrangian}  on a (bosonic) hypersurface embedded in superspace, defined as the Deformed coset $\rm \rm\tilde G/H$.\footnote{Indeed in most supergravity {theories} we may have a larger group of gauge invariance other than the Lorentz one.  When this happens{,} and we want to start from the full supergroup manifold, the factorization of the extra coordinates belonging to ${\rm H}^\prime$ can be obtained from the action principle exactly as for the Lorentz group coordinates.} Alternatively{,} we could start integrating on the whole $\rm\tilde G$ and obtaining the factorization of the coordinates of  ${\rm H}$ as {field equations}. Since factorization of the gauge group $\rm H$ is actually always true if the {Lagrangian} is $\rm H${-}invariant,  the customary way to proceed is to start with a $\rm H$ invariant {Lagrangian} on  superspace $\rm\tilde G/ \rm H$.
\item
The field equations must reduce to identities if all the curvatures are zero, that is if we are in the vacuum configuration with left-invariant1-forms $\mu^A=\sigma^A$.
\end{itemize}

The field equations derived from such {Lagrangian} give equations of two types:

\begin{enumerate}[i]
\item Field equations relating \emph{outer} components of the  curvatures linearly in terms of the \emph{inner} ones{.} These are the rheonomic conditions equivalent to saying that the {Lagrangian} is supersymmetric. Indeed as we have seen supersymmetry is {an} invariance of the {Lagrangian} under diffeomorphisms in superspace and the the \emph{rhenomic} relations simply express the fact that the theory is independent {of} the space-time identification of the hypersurface  $\mathcal M_D$ embedded {in} superspace. Note that in solving these equations all undetermined coefficients of the various terms become fixed.
\item Field equations which are differential equations in superspace. Restricted to the bosonic hypersurface $\mathcal M_D$ they are the space-time equations of motion.
\end{enumerate}

\subsection{{The Role of the Bianchi Identities.}}

Till now we have  extensively explained how to give a geometrical interpretation to the supersymmetry transformations of supergravity  either  constructed on superspace, when factorization of the Lorentz coordinates is assumed a priori, or directly starting from the group manifold locally identified by the underlying supersymmetry algebra.\\
 There is however an equivalent and powerful approach to the construction of the equations of motion and transformation laws which is based on a systematic use of the Bianchi identities \emph{ assuming rheonomy} from the very beginning.

 To understand this point we note that the Bianchi identities are true identities only if no constraint is assumed among the supervielbein components of the curvatures. However what rheonomy does is exactly to give relations among the outer {components $R^A_{B,\alpha}$} ({those} along at least one fermionic vielbein $\psi$) and  inner components $R^A_{ab}$ (namely the components on the cotangent space to space-time). Moreover, in almost any case, one also assumes a further constraint called kinematical constraint, namely vanishing super-torsion $\hat T^a=0$. The Bianchi identities then assume the form of differential constraints among the space-time components. These differential constraints, on the other hand, can be nothing else than the equations of motion, since Bianchi identities cannot conflict with the differential equations obtained from the Lagrangian. Once the field equations are obtained{,} the {Lagrangian}, if desired, can be easily reconstructed.
 In the actual computations one usually couples the two methods, namely the Lagrangian approach and the Bianchi identities equations, to arrive in the simplest way to the final determination of the parametrization of the curvatures in superspace (and thus to the supersymmetry transformation laws) and to the determination of all the coefficients in the {Lagrangian}.

\section{Results of the Geometric Approach. Some Remarks.}

Most of the previous considerations have been dedicated to the geometrical interpretation of supersymmetry and to the explicit geometric construction of a supergravity theory.
 A natural question is now what has been the impact of this {kind} of approach from the physical point of view. We cannot of course enter in a detailed exposition of the results obtained from the very beginning till nowadays. We limit ourselves to give some remarks and observations concerning its power in treating  higher dimensional supergravities and matter coupled theories  containing antisymmetric tensor fields, together with some unexpected  properties of superspace.  As far as the most relevant  results obtained in the geometric approach is concerned, we limit ourself to give a short list of them in the Appendix. Here is  a couple of short and hopefully interesting comments:\\.
\begin{itemize}
\item Among the  most interesting results there is certainly the reduction of  theories containing antisymmetric {tensors} to equivalent theories formulated in terms of  super-Lie {algebras}. This result was obtained in reference {\cite{D'Auria:1982nx}} from the analysis of the geometrical formulation of {$D=11$} supergravity formulated on space-time in reference {\cite{Cremmer:1978km}}.
 As {$D=11$} supergravity is thought to be the low energy limit of the so-called $M$-theory, this result can have a special relevance for a better understanding of the group theoretical structure of $M$-theory. Because of its importance, we shall devote the next section to a short review of this approach in the case of $D=11$ supergravity.
 \item It is undeniable that the systematic use of the geometric and group{-}theoretical approach has been an essential tool to obtain many other interesting results. For example, we can say that, in general, the introduction of matter coupling to {pure} supergravities allows to put in light the global and local symmetries inherent non-linear interaction structure of the coupling to matter multiplets. Indeed very often the use of the geometrical approach has allowed to arrive to a complete {answer} to problems where other approaches often had given only limited answers.\footnote{Most of these developments and results can be found in the excellent review  \cite{Trigiante:2016mnt}.}

 A typical example was the full construction of the ${\mathcal{N}}=2$, $D=4$ matter coupled supergravity {\cite{Castellani:1991et, Andrianopoli:1996cm, DAuria:1990qxt}}, which was previously formulated using the superconformal approach in a coordinate dependent way \cite{deWit:1984rvr}. The geometrical approach provides a complete Lagrangian and transformation laws quite independently of the coordinate used for the  scalar fields description of the $\sigma$ model and  it makes the introduction of the notion of Special $\rm\tilde G$eometry, the geometry of the  Special and Quaternionic manifolds, the momentum maps and the related  gaugings, together with a complete description of the scalar potential, very natural. These results also give insight into the related superconformal two{-}dimensional theories and Calabi{-}Yau compactifications in string theory. Further developments are also to be found in reference \cite{Trigiante:2016mnt}.
 \item Another interesting observation is the following: The geometric approach discussed in this pedagogical review is naturally formulated in superspace. One could ask wether this approach is exactly equivalent to the purely space-time approach. This seems not be the case in some theories, like $\mathcal{N}=1$, $D=6$ {\cite{DAuria:1983jkr}}, and $D=10$, $IIB$ \cite{Castellani:1993ye}.\\ In the $D=6$  supergravity the field multiplet contains the sechsbein{,} a Weyl gravitino{,} and a 2-form (that is an antisymmetric two-index tensor), it was shown  that a consistent theory on the {group manifold} might have no counterpart in the usual Noether approach. In our geometric $D=6$ superspace model, the self-duality of the 2-form field-strength, necessary to match the Bose-Fermi on-shell degrees of freedom, follows from {group manifold/superspace} variational equations, but not from their $x$-space restriction. As a consequence, the theory is consistent, although the $x$-space {Lagrangian} is not supersymmetry invariant. Exactly in the same way can be treated the $D=10$, $IIB$ theory, so that also in this case the self-duality of the 5-form can be retrieved from the superspace equations of motion. These results hint to extra properties of  superspace which can be traded on the embedded hypersurface $\mathcal M_4$ only after the superspace equations of motion have been implemented, that is they are not visible using a purely space-time approach.
\end{itemize}

\section{Higher p-Forms Supergravities and their Hidden Supergroups.}

We have often stressed that the mechanism of rheonomy actually holds in all supergravities, independently of the number of supersymmetries, the dimensionality of space-time{,} and their matter couplings, if any. However, apart from few exceptions, most of the higher dimensionality theories have a gravitational multiplet containing antisymmetric tensors of higher rank, mostly of rank two. Similarly,  matter  supermultiplets also can have higher rank tensors. In these cases the group manifold interpretation  presented before as a possible starting point for supergravities whose fields are defined on a group manifold cannot be maintained. Indeed the coadjoint multiplet of a (super{-}) group consists of 1-forms dual to the group generators, with no room for higher p-forms. \\
In the present section we will show, referring mainly to the case of {$D=11$} supergravity where this development was first presented \cite{D'Auria:1982nx}, {that}:
\begin{itemize}
\item The Maurer-Cartan equations can be generalized to more general structures, called Free Differential Algebras {(FDAs)}, admitting in their multiplet also forms of {degree higher than one}. They represent a natural extension of {Lie algebras} in their dual formulation and can accommodate  supermultiplets containing higher p-forms satisfying the integrability requirement $d^2=0$.
\item Each higher {p-form} $A^{(p)}$ can be decomposed  in  terms of a set of trilinear (wedge) products of $p$ 1-forms, where besides the supervielbein basis {$(V^a\,,\psi)$} there appear new 1-forms  valued in tensor or spinor representations of the Lorentz group. The new 1-forms obey extra  Maurer-Cartan equations, besides those  of the super{-}Poincar\'e group. The decomposition can be  done in such a way that the coefficients of the polynomial written as a  sum of products of $p$ 1-forms assure the integrability of the original FDA equation for the p-form $A^{(p)}$.
\item Together with the super-Poincar\'e dual generators, the new Maurer-Cartan {equations} describe the dual form of  a new super-Lie algebra which, at least locally{,} describes a group manifold called the \emph{{h}idden supergroup} of the FDA which in a sense can be considered as the group-theoretical starting point for a construction of the supergravity theories possessing higher p-forms in their gravitational multiplet.
\item Among the new 1-forms needed to assure that the given decomposition reproduces  the integrable equation of the FDA,  there appear  extra spinor 1-forms (one in the case of $D=11$ supergravity) whose dual generators $Q_\alpha^\prime$ in the  Lie superalgebra are nilpotent. Their role, as recently clarified \cite{Andrianopoli:2016osu, Andrianopoli:2017itj}{,} is to assure that the new 1-form fields thus introduced are gauge fields living on the fiber bundle whose base space is ordinary superspace, so that their dependence on the new coordinates are completely factorized. This means that their curvatures are horizontal and do not add new degrees of freedom other than those already present in the original FDA.
This property  works exactly in the same way for all higher dimensional {theories} with $D < 11$ whose gravitational multiplet contain antisymmetric tensor fields, for example in $\mathcal{N}=2$, $D=7$ supergravity where two such nilpotent spinor generators are present.
\end{itemize}

These results were obtained in \cite{D'Auria:1982nx} {by R. D'Auria and P. Fr\'e} in the case of the maximal $D=11$ supergravity trying to give a fully geometrical interpretation of the space-time formulation of the theory \cite{Cremmer:1978km}. Indeed, in this theory there {appears} an  antisymmetric tensor of rank three in the gravitational multiplet. In their approach{,} R. D'Auria and P. Fr\'e introduced for the first time the generalization of the Maurer-Cartan equations for integrable systems containing higher p-forms which they called Cartan Integrable Systems (CIS).  Only later it was realized that structures of this kind were already introduced in mathematics and called \emph{Free Differential Algebras} {\cite{Sullivan}}, which is the name now universally accepted. \\
Even if {we} shall not give any account of the underlying mathematics, it must be said that the relation between the FDA and groups or supergroups relies on the \emph{Chevalley-Eilenberg (super)-Lie algebra cohomology groups}. The procedure {we previously alluded to of} decomposing a higher {p-form} in a polynomial of Lorentz valued 1-forms is the inverse of the construction of {a} FDA starting from a (super{-})Lie algebra, which, as far as I know, {was} not treated in the Lie algebra cohomology theory.

In the following{,} we give a short account of the FDA of $D=11$ supergravity and its resolution as {a} \emph{hidden} ordinary Lie supergroup.

\section{Free Differential Algebra and Hidden Supergroup of {$D=11$} Supergravity.}

{Eleven-dimensional supergravity} can be founded on the following FDA:
\begin{eqnarray}
R^{ab}&\equiv& d\omega^{ab} - {\omega^{ac}\wedge \omega_c^{\;b}}=0\,,\label{FDA11omega}\\
T^a&\equiv& D V^a - \frac{{\ii}}{2}\overline{\Psi}\wedge \Gamma^a \Psi =0\,,\label{FDA11v} \\
\rho &\equiv &D\Psi=0\,,\label{FDA11psi}\\
F^{(4)} &\equiv& dA^{(3)} - \frac{1}{2}\overline{\Psi}\wedge \Gamma_{ab}\Psi \wedge V^a \wedge V^b =0\,,\label{FDA11a3} \\
F^{(7)}&\equiv& dB^{(6)} - 15 A^{(3)}\wedge  dA^{(3)} -\frac{{\ii}}{2}\overline{\Psi}\wedge \Gamma_{a_1\ldots a_5}\Psi \wedge V^{a_1} \wedge \cdots {\wedge} V^{a_5}=0\,.\label{FDA11b6}
\end{eqnarray}
Note that this differential system is an extension of the usual Maurer-Cartan {equations} and as such it only describes the structure of the physical vacuum of the theory.\\
In our case the super-Poincar\'e Maurer-Cartan equations in $D=11$ with the addition of the higher order differential of the 3-form $A^{(3)}$ and 6-form $B^{(6)}${.}\footnote{In the original paper {the} last equation (\ref{FDA11b6}) was not present. Actually, it was almost immediately realized (see e.g. reference \cite{Castellani:1991et}, {Vol. 2}) that, besides  the simplest FDA including as exterior form only $A^{(3)}$,  one can  extend the FDA to include also a (\emph{magnetic}) 6-form potential $B^{(6)}$, related   to $A^{(3)}$ by  Hodge-duality of the corresponding field-strengths on space-time. Indeed there is a constructive procedure based on the Chevalley-Eilenberg Lie {a}lgebra {c}ohomology to arrive to the maximal extension of the super-Poincar\'e algebra given by the FDA  which implies as a maximal extension the presence of the $B^{(6)}$ \cite{Castellani:1991et}.} \\
The consistency of the FDA requires the integrability of the last two equations, $ d^2 A^{(3)}=0$, $ d^2 B^{(6)}=0$.  It can be shown  that {this is} in fact satisfied as a consequence of 3-fermion Fierz identities obeyed by the gravitino {1-form field in} eleven dimensions (see e.g. reference \cite{Castellani:1991et}){.}\footnote{Note that here and in the following we do not elaborate on the theory out of vacuum, namely the  interacting theory, since the topological {(and cohomological)} structure of the theory, which will be the object of the present investigation, is fully {caught} by the ground state of the  FDA.}

The authors of \cite{D'Auria:1982nx} asked themselves whether one could trade the FDA structure on which the theory is based with an ordinary   Lie superalgebra, written in its dual Cartan form, that is in terms of 1-form gauge fields which turn out to be valued in non{-}trivial tensor and spinor representations of  Lorentz group $\rm{SO}(1,10)$. This would {allow to} disclose the fully extended Lie superalgebra hidden in the supersymmetric FDA. {This was proven to be true, and the hidden superalgebra underlying the FDA describing $D=11$ supergravity was presented for the first time.

 It was shown that this is indeed possible by associating, to the  forms $A^{(3)}$ and $B^{(6)}$, the bosonic 1-forms $B_{ab}$ and $B_{a_1 \cdots a_5}$,  in the antisymmetric representations of $\rm{SO}(1,10)$, and furthermore  an extra \emph{spinor} 1-form $\eta$. The Maurer-Cartan equations satisfied by these new 1-forms are:
 \begin{eqnarray}
 \mathcal D B_{a_1a_2} & = & \frac{1}{2}\overline{\Psi}\wedge\Gamma_{a_1a_2}\Psi , \\
\label{also} \mathcal D B_{a_1...a_5}& = & \frac{i}{2} \overline{\Psi}\wedge \Gamma_{a_1...a5}\Psi\\
 \mathcal D \eta & = & i E_1 \Gamma_a \Psi \wedge V^a + E_2 \Gamma^{ab}\Psi \wedge B_{ab}+ i E_3 \Gamma^{a_1...a_5}\Psi \wedge B_{a_1...a_5}\,, \label{Deta}
\end{eqnarray}
$\mathcal D$ being the  Lorentz-covariant derivatives.
Of course the whole consistence of this approach also requires the $d^2$ closure of the Maurer-Cartan newly introduced fields  $B_{ab}$, $B_{a_1\cdots a_5}$ and $\eta$. For the two bosonic 1-form fields the  $d^2$ closure is obvious in the ground state, because of the vanishing of the curvatures $R^{ab}$ and $\rho=\mathcal D \psi$, while $\mathcal D^2\eta=0$  requires the further condition:
\begin{equation}\label{integrability11}
E_1+10E_2-720E_3=0\, ,
\end{equation}
which can be derived using the Fierz identities of the wedge product of three gravitino 1-forms in superspace \cite{D'Auria:1982nx}.\\
 In reference \cite{D'Auria:1982nx} the   most general decomposition of the 3-form $A^{(3)}$ in terms of the  1-forms $B_{ab}$,  $B_{a_1...a_5}$ and $\eta$  was presented. It has the following form:
\begin{eqnarray}\label{29}
A^{(3)} & = & T_0 B_{ab} \wedge V^a \wedge V^b + T_1 B_{a b}\wedge B^{b} _{\;c}\wedge B^{c a}+ \nonumber\\
& +& T_2 B_{b_1 a_1...a_4}\wedge B^{b_1}_{\; b_2}\wedge B^{b_2 a_1...a_4}+ T_3 \epsilon_{a_1...a_5 b_1...b_5 m}B^{a_1...a_5}\wedge B^{b_1...b_5}\wedge V^m + \nonumber\\
& +& T_4 \epsilon_{m_1...m_6 n_1...n_5}B^{m_1m_2m_3p_1p_2}\wedge B^{m_4m_5m_6p_1p_2}\wedge B^{n_1...n_5} + \nonumber\\
& + & i S_1 \overline{\Psi}\Gamma_a \eta \wedge V^a + S_2 \overline{\Psi}\Gamma^{ab}\eta \wedge B_{ab}+ i S_3 \overline{\Psi}\Gamma^{a_1...a_5}\eta \wedge B_{a_1...a_5}\,,\label{a3par}
\end{eqnarray}
where  $\rm T_i$ and $\rm S_j$ are numerical coefficients.
To show the equivalence of the FDA with a ordinary super-Lie algebra (in dual form) it is required that
  the integrability condition in superspace of the 3-form, $d A^{(3)}$, computed in terms of differentials of the new 1-forms gives the same results as in the case of equation (\ref{FDA11a3}), namely  $ dA^{(3)}-\frac{1}{2}\overline{\Psi}\wedge \$\rm\tilde Gamma_{ab}\Psi \wedge V^a \wedge V^b =0$. To obtain such integrability the extra terms containing the currents involving the \emph{extra spinor 1-form $\eta$ turn out to be necessary.} The Ansatz (\ref{a3par}) restricted to the bosonic 1-forms does not work. In other words the inclusion of the spinor 1-form field $\eta$ enters in the decomposition of the 3-form $A^{(3)}$ in such a way to properly reproduce the vacuum FDA on ordinary superspace.

   When the integrability is implemented  all the coefficients in the decomposition become  fixed in terms of the ratio $ E_3/E_2$.\footnote{In \cite{D'Auria:1982nx} the first coefficient $T_0$ was arbitrarily fixed to $T_0=1$ giving only 2 possible solutions for the set of parameters $\{T_i,S_j,E_k\}$. It was pointed out later in \cite{Bandos:2004xw, Bandos:2004ym} that this restriction can be relaxed thus giving a more general solution in terms of one parameter. Indeed, as  observed in the quoted reference, one of the $E_i$ can be reabsorbed in the normalization of $\eta$, so that, owing to the relation (\ref{Deta}), we are left with one free parameter, say $E_3/E_2$.}
In this way{,} one arrives at the following  full set of Maurer-Cartan equations for the left-invariant 1-forms  $(\omega^{ab}\,, V^a\,,\psi\,, B_{a b}\,, B_{a_1 \ldots a_5}\,,\eta)${:}
\begin{eqnarray}
 \label{deta1} d\omega^{ab}&=& \omega^{ac}\wedge \omega_c^{\;b} ,\\
  \label{deta2}D V^a &=& \frac{\ii}{2}\overline{\Psi}\wedge \Gamma^a \Psi, \\
 \label{deta3}D \Psi&=&0, \\
  \label{deta4}D B_{a_1a_2} & = & \frac{1}{2}\overline{\Psi}\wedge \Gamma_{a_1a_2}\Psi , \\
 \label{deta5}D B_{a_1 \ldots a_5}& = & \frac{\ii}{2} \overline{\Psi}\wedge \Gamma_{a_1 \ldots a_5}\Psi,\\
 \label{deta6} D \eta & = & \ii E_1 \Gamma_a \Psi \wedge V^a + E_2 \Gamma^{ab}\Psi \wedge B_{ab}+ \ii E_3 \Gamma^{a_1 \ldots a_5}\Psi \wedge B_{a_1 \ldots a_5}\,.
\end{eqnarray}
{This set of Maurer-Cartan equations identifies the supergroup which is (locally) described, as anticipated, by the hidden super-Lie algebra underlying the theory (this same superalgebra written in terms of commutators is given below).} \\
It must be noted the if we neglect the presence of the $\eta$ spinor 1-form in the decomposition (\ref{29}) we can obtain a closed algebra which, however, is not equivalent to the FDA we started from since it fails to give the closure $d^2 A^{(3)}=0$. Only when the extra currents containing $\eta$  are present in (\ref{29}) we obtain such integrability.\\
Let us finally write down the hidden superalgebra in terms of generators closing a set of (anti)commutation relations.

{To recover the superalgebra  in terms of (anti)commutators of the dual Lie superalgebra generators}
\begin{equation}
T_{{A}} \equiv\{P_a, Q, J_{ab},  Z^{ab}, Z^{a_1 \ldots a_5}, Q'\}\,,\label{t11d}
\end{equation}
which are dual to the 1-forms $\left(V^a,\,\psi,\,\omega^{ab}\,,B_{ab},\, B_{a_1 \ldots a_5},\,\eta\right)$ respectively, one uses the duality between 1-forms and generators and  finds that the $D=11$ FDA  corresponds to the following hidden  superalgebra (besides the Poincar\'e algebra):
\begin{eqnarray}
\left \{Q,\bar Q\right\} &=& -\left({\ii} \Gamma^a P_a + \frac12 \Gamma^{ab}Z_{ab}+ \frac {{\ii}}{5!} \Gamma^{a_1 \ldots a_5}\,Z_{a_1 \ldots a_5}\right)\,, \label{qq11}\\ \nonumber
\left[ Q',\bar Q' \right] &=& 0\,,\\ \nonumber
[Q, P_a] &=& -2 {\ii} E_1 \Gamma_a Q'\,,\\ \nonumber
[Q, Z^{ab}] &=&-4 E_2 \Gamma^{ab}Q' \,, \\ \nonumber
[Q, Z^{a_1 \ldots a_5}] &=&- 2 \,(5!) {\ii} E_3 \Gamma^{a_1 \ldots a_5}Q'\,, \\ \nonumber
[J_{ab}, Z^{cd}]&=&-8 \delta^{[c}_{[a}Z_{b]}^{\ d]}\,,\\ \nonumber
[J_{ab}, Z^{c_1\dots c_5}]&=&- 20 \delta^{[c_1}_{[a}Z^{c_2\dots c_5]}_{b]}\,,\\ \nonumber
[J_{ab}, Q]&=&- \Gamma_{ab} Q\,,\\ \nonumber
 [J_{ab}, Q']&=&- \Gamma_{ab} Q'\,.\label{SLA}
\end{eqnarray}
All the other commutators (beyond the Poincar\'e part) vanish.  We shall identify the super-Lie algebra in either of the two dual forms given as {M}aurer-Cartan equations or (anti-commutators)  as D'Auria-Fr\'e algebra (DF-algebra in the following).
  In the Lie algebra version of the dual Maurer-Cartan equations the  1-form{s}  {$B_{a_1\ldots a_5}$ and  $B_{ab}$} are the 1-forms dual to the central generators {$Z^{a_1 \ldots a_5}$ and $Z^{ab}$, respectively,} of a central extension of the supersymmetry algebra given by the usual $D=11$ super-Poincar\'e algebra and  including the extra nilpotent generator $Q^\prime$ {(dual to the spinor 1-form $\eta$)}.\footnote{Here and in the following the term ``central'' for the charges {$Z^{ab}$, $Z^{a_1\ldots a_5}$, and for the spinorial charge $Q'$} refers to their commutators with all the generators apart from the Lorentz generator $J_{ab}$. The commutation relations with $J_{ab}$ are obviously dictated by their Lorentz index structure.} Actually, as shown in reference \cite{Andrianopoli:2017itj}, from a cohomological point of view, to reproduce the integrability of $dA^{(3)}$,  the presence of the 1-form $B_{a_1...a_5}$ in the decomposition (\ref{29})  is not necessary, since all the terms where it appears sum up to give an exact 3-form. However, as we have seen, even if cohomologically trivial, its addition allows to extend the Lie superalgebra in a non-trivial way.\footnote{More precisely in reference \cite{Andrianopoli:2017itj} it was shown that once formulated in terms of its hidden superalgebra of 1-forms, $A^{(3)}$ can be actually decomposed into the sum of two parts having different group-theoretical meaning: One of them does not depend on $B_{a_1 \ldots a_5}$ and allows to reproduce the FDA describing $D=11$ supergravity,  while the second one does not contribute to the 4-form cohomology, being a closed 3-form in the vacuum; however, the second part defines a one parameter family of trilinear forms invariant under a symmetry algebra that is related to $\mathfrak{osp}(1|32)$ by redefining the spin connection and adding a new Maurer–Cartan equation.  Correspondingly, also the spinor 1-form $\eta$ can be analogously split into two different spinors appearing each one in just one of the two parts in which $A^{(3)}$ is decomposed.}

\subsection{D=11 Supergravity and M-theory.}

After several years, on the basis of different considerations, the same algebra, but \emph{without the inclusion of the extra nilpotent generator} was rediscovered. This superalgebra, actually a subalgebra of the hidden superalgebra (\ref{qq11}),  was named $M$-algebra \cite{deAzcarraga:1989mza, Sezgin:1996cj, Townsend:1997wg, Hassaine:2003vq, Hassaine:2004pp}.

 The presence  in the relation{s} (\ref{deta1})-(\ref{deta6}) of the bosonic hidden 1-forms $B_{ab}, B_{a_1 \ldots a_5}$ can be considered as a  generalization of the centrally extended supersymmetry algebra of \cite{Witten:1978mh} (where the central generators were  associated with electric and magnetic charges), and{,} as such, have in fact a topological meaning.  This  was recognized in {\cite{deAzcarraga:1989mza} and \cite{Achucarro:1987nc}},  where it was shown they are to be associated with extended objects  (2-brane and 5-brane {charges, respectively}) in space-time. {The $M$-algebra} is commonly considered as the super{-}Lie algebra underlying $M$-theory \cite{Schwarz:1995jq, Duff:1996aw, Townsend:1996xj} in its low energy limit, corresponding to {$D=11$} supergravity in the presence of non-trivial $M$-brane sources \cite{Achucarro:1987nc, Bergshoeff:1987cm, Duff:1987bx, Bergshoeff:1987qx, Townsend:1995kk, Townsend:1995gp}.

 A field theory based on the {$M$-algebra,} however, is naturally described on the \emph{enlarged superspace} whose cotangent space is spanned, besides the gravitino 1-form, also by bosonic fields $\{V^a, {B_{ab}, B_{a_1\ldots a_5}} \}$. If we hold on the idea that the low energy limit of $M$-theory should be based on the same ordinary superspace, spanned by the supervielbein $(V^a,\psi)$, as in the original formulation of {$D=11$} supergravity \cite{Cremmer:1978km},  then the $M$-algebra cannot be the final answer since it does not contain the extra 1-form $\eta$ dual to the nilpotent generator  fermionic generator $Q'$. Indeed we have shown that in order to reproduce the FDA on which {$D=11$} supergravity is based the presence of $\eta$ among the 1-form generators is necessary. Actually the DF-{algebra}, (\ref{deta1})-(\ref{deta6}) or (\ref{qq11}),
contains the $M$-algebra as a subalgebra since it  also \emph{includes  a nilpotent {\rm($Q'^2=0$\rm)} fermionic generator $Q'$} dual to the spinor 1-form $\eta$ whose contribution to the Maurer-Cartan equations of {the} DF-algebra is given by equation {(\ref{deta6})}. \\ In other words the DF-algebra underlying the formulation of the {eleven-dimensional} FDA on superspace reproduces the {eleven-dimensional} theory on space-time introduced in reference \cite{Cremmer:1978km}}, if and only if the decomposition of the 3-form $A^{(3)}$ \emph{also includes the {1-form $\eta$}}.\\
As it was shown in reference \cite{Andrianopoli:2016osu},  this  in turn implies that the group manifold generated by the DF-{algebra} has a fiber bundle structure whose base space is ordinary superspace, while the fiber is spanned, besides the Lorentz spin connection $\omega^{ab}$, also by the bosonic 1-form generators {$B_{ab}, B_{a_1,\dots a_5}$}. It follows that if we would start from the group manifold generated by the DF{-}algebra, the coadjoint multiplet would now be {$\mu^A=(\omega^{ab}, B_{ab},B_{a_1 \dots a_5},V^a,\psi, \eta)$} and the action principle would factorize the coordinates associated to the first three 1-forms so that their degrees of freedom would not enter into the equations of motion. Indeed the presence of $Q'$, dual to the 1-form $\eta$, allows to consider the extra 1-forms {$B_{ab}$} and {$B_{a_1\ldots a_5}$} as gauge fields in ordinary superspace instead of  additional vielbeins of an enlarged superspace, that is, their curvatures on the fiber are \emph{horizontal}. This is due to the dynamical cancellation of their unphysical contributions to the supersymmetry and gauge transformations with the supersymmetry and {gauge transformations of $\eta$.\footnote{As observed in \cite{Andrianopoli:2016osu}, all the above procedure of enlarging the field space to recover a well defined description of the physical degrees of freedom is strongly reminiscent of the BRST-procedure, and the behavior of $\eta$ is such that it can be actually thought of as a ghost for the 3-form gauge symmetry, when the 3-form is parametrized in terms of 1-forms.}

\section{Conclusions.}

We have shown how the original idea formulated by Y. Ne'eman and T. Regge of defining gravity and supergravity theories directly on a (super-)group  manifold $\rm\tilde G$ actually allows a completely geometrical and group-theoretical formulations of any gravity or supergravity extended theory.

We have explained how the generalized action principle of integrating the Lagrangian on a submanifold of $\rm\tilde G$ is capable of obtaining the following results:

Factorization of coordinates belonging to gauge subgroups of $\rm\tilde G$ can be obtained by using a generalized action principle where space-time is represented by a bosonic submanifold immersed in the (super-)group, so that $\rm\tilde G$ actually becomes endowed by a fiber bundle structure.

For supergroups, where the notion of superspace as base space of the fiber bundle appears,  the same procedure based on the extended action principle allows to add, to the notion of factorization, the notion of supersymmetry whose geometrical meaning is that the field-strengths (curvatures) in superspace are not horizontal, but can be expressed linearly in terms of the space time field-strengths. The corresponding geometrical interpretation has been named \emph{rheonomy}.

 It is possible to give very simple \emph{building principles} that allow an almost algorithmic procedure for the construction of any supergravity Lagrangian, the most important principle being \emph{geometricity}.

 Besides the original results obtained not only in supergravity, but also in related questions like duality, Calabi-Yau compactifications, mononodromies etc, an important step has been the discovery of the Free Differential Algebras as a geometrical way to formulate  supergravities containing higher p-forms fields by a suitable generalization of  the Maurer-Cartan equations to higher p-forms. In particular using the notion of Lie algebra cohomology  it has been possible to show that any such supergravity containing higher p-forms can be reduced to an ordinary supergravity based on an ordinary Lie superalgebra.

\section{Acknowledgements}

 I thank my friends and colleagues L. Andrianopoli, L. Ravera and M. Trigiante for their critical reading of the manuscript and interesting suggestions.

\section{Appendix}

We give a short list of the most relevant and interesting results obtained in the geometric approach.

Besides giving the geometrical structure of the already existing supergravity theories, some supergravity theories were first obtained in the geometric approach, see e.g. ${\mathcal{N}}=3$, $D=4$ and $\mathcal{N}=2$, $D=6$, $\rm F_4$ matter coupled  supergravities  \cite{Castellani:1985ka, DAuria:2000afl}. \\
Construction of central and matter charges and symplectic structure of all {$D= 4$}, ${\mathcal{N}}$–extended theories {\cite{Andrianopoli:1996ve}}. \\
Symplectic invariant coupling of scalar-tensors \cite{Trigiante:2016mnt, DallAgata:2003sjo, DAuria:2004yjt} and vector-tensor multiplets \cite{Andrianopoli:2011zj} in ${\mathcal{N}}=2$ supergravity and the role of magnetic charges.\\
Derivation of {the} ${\mathcal{N}}=1$ and ${\mathcal{N}}=2$, $D=4$ supergravity {Lagrangians} in the presence of a boundary \cite{Andrianopoli:2014aqa}. \\
Unexpected interesting relation between ${\mathcal{N}}=2$ supergravity in $D=4$ and a three{-}dimensional theory describing the graphene electronic properties \cite{Andrianopoli:2018ymh}.\\
Other relevant results are the following:
Anomaly Free supergravity in $D=10$ \cite{DAuria:1987tdr}; Duality transformations in supersymmetric Yang-Mills theories coupled to supergravity {\cite{Ceresole:1995jg}}; A detailed analysis of the role played by the Picard-Fuchs equations in supergravity {\cite{Ceresole:1992su, Ceresole:1993qq, Billo:1995ge}}; Symplectic structure of ${\mathcal{N}}=2$ supergravity and its central extension \cite{Ceresole:1995ca}; Symplectic structure and monodromy group for the Calabi-Yau two moduli space \cite{Ceresole:1993nz, Cadavid:1995bk} (as far as rigid supersymmetry is concerned, the necessity of introducing a Chern-Simon term in $D=10$ super Yang-Mills  ${\mathcal{N}}=1$ Lagrangian was first  realized using the geometric approach, see reference {\cite{DAuria:1981zjr}}). \\
Finally, even if not concerning supersymmetry, an important result has been obtained in the theory of gravitation using a completely geometric {Lagrangian} coupled to a pseudo-scalar field in a non-canonical way. Using this {Lagrangian} it has been possible to show the existence of \emph{symptotically flat gravitational instantons in gravity} \cite{DAuria:1981ddz}. Further results in this approach were also obtained in reference \cite{Chandia:1997hu}.


\begin{thebibliography}{99}

\bibitem{Neeman:1978njh}
  Y.~Ne'eman and T.~Regge,
  ``Gauge Theory of Gravity and Supergravity on a Group Manifold,''
  Riv.\ Nuovo Cim.\  {\bf 1N5} (1978) 1,  and
 Phys.Lett. 74B (1978) 54-56;

\bibitem{D'Auria:1982nx}
  R.~D'Auria and P.~Fr\'{e},
  ``Geometric Supergravity in d = 11 and Its Hidden Supergroup,''
  Nucl.\ Phys.\ B {\bf 201} (1982) 101
   Erratum: [Nucl.\ Phys.\ B {\bf 206} (1982) 496].

\bibitem{Sati:2015yda}
  H.~Sati and U.~Schreiber,
  ``Lie n-algebras of BPS charges,''
  JHEP {\bf 1703} (2017) 087
  doi:10.1007/JHEP03(2017)087
  [arXiv:1507.08692 [math-ph]].

\bibitem{stronghom}
https://ncatlab.org/nlab/show/L-infinity+algebras+in+physics

\bibitem{Castellani:1991et}
  L.~Castellani, R.~D'Auria and P.~Fr\'e,
  ``Supergravity and superstrings: A Geometric perspective,'' Vol. 1 and 2.
 Singapore: World Scientific (1991)

\bibitem{Cremmer:1978km}
  E.~Cremmer, B.~Julia and J.~Scherk,
  ``Supergravity Theory in Eleven-Dimensions,''
  Phys.\ Lett.\ B {\bf 76} (1978) 409
   [Phys.\ Lett.\  {\bf 76B} (1978) 409].

\bibitem{Trigiante:2016mnt}
  M.~Trigiante,
  ``Gauged Supergravities,''
  Phys.\ Rept.\  {\bf 680} (2017) 1
  doi:10.1016/j.physrep.2017.03.001
  [arXiv:1609.09745 [hep-th]]

\bibitem{Andrianopoli:1996cm}
  L.~Andrianopoli, M.~Bertolini, A.~Ceresole, R.~D'Auria, S.~Ferrara, P.~Fr\'{e} and T.~Magri,
  ``N=2 supergravity and N=2 superYang-Mills theory on general scalar manifolds: Symplectic covariance, gaugings and the momentum map,''
  J.\ Geom.\ Phys.\  {\bf 23} (1997) 111
  doi:10.1016/S0393-0440(97)00002-8
  [hep-th/9605032].

\bibitem{DAuria:1990qxt}
  R.~D'Auria, S.~Ferrara and P.~Fr\'{e},
  ``Special and quaternionic isometries: General couplings in N=2 supergravity and the scalar potential,''
  Nucl.\ Phys.\ B {\bf 359} (1991) 705.
  doi:10.1016/0550-3213(91)90077-B

\bibitem{deWit:1984rvr}
  B.~de Wit, P.~G.~Lauwers and A.~Van Proeyen,
  ``Lagrangians of N=2 Supergravity - Matter Systems,''
  Nucl.\ Phys.\ B {\bf 255} (1985) 569.
  doi:10.1016/0550-3213(85)90154-3

\bibitem{DAuria:1983jkr}
  R.~D'Auria, P.~Fr\'{e} and T.~Regge,
  ``Consistent Supergravity in Six-dimensions Without Action Invariance,''
  Phys.\ Lett.\  {\bf 128B} (1983) 44.
  doi:10.1016/0370-2693(83)90070-9

\bibitem{Castellani:1993ye}
  L.~Castellani and I.~Pesando,
  ``The Complete superspace action of chiral D = 10, N=2 supergravity,''
  Int.\ J.\ Mod.\ Phys.\ A {\bf 8} (1993) 1125.

\bibitem{Andrianopoli:2016osu}
  L.~Andrianopoli, R.~D'Auria and L.~Ravera,
  ``Hidden Gauge Structure of Supersymmetric Free Differential Algebras,''
  JHEP {\bf 1608} (2016) 095.
  doi:10.1007/JHEP08(2016)095
  [arXiv:1606.07328 [hep-th]].

\bibitem{Andrianopoli:2017itj}
  {L.~Andrianopoli, R.~D'Auria and L.~Ravera,
  ``More on the Hidden Symmetries of 11D Supergravity,''
  Phys.\ Lett.\ B {\bf 772} (2017) 578
  doi:10.1016/j.physletb.2017.07.016
  [arXiv:1705.06251 [hep-th]].}

\bibitem{Sullivan}
 D.~Sullivan,
 ``Infinitesimal computations in topology". Publications Math\'ematiques de l'IHES, 47 (1977), p. 269-331

\bibitem{Bandos:2004xw}
  I.~A.~Bandos, J.~A.~de Azcarraga, J.~M.~Izquierdo, M.~Picon and O.~Varela,
  ``On the underlying gauge group structure of D=11 supergravity,''
  Phys.\ Lett.\ B {\bf 596} (2004) 145
  doi:10.1016/j.physletb.2004.06.079
  [hep-th/0406020].

\bibitem{Bandos:2004ym}
  I.~A.~Bandos, J.~A.~de Azcarraga, M.~Picon and O.~Varela,
  ``On the formulation of D = 11 supergravity and the composite nature of its three-form gauge field,''
  Annals Phys.\  {\bf 317} (2005) 238
  doi:10.1016/j.aop.2004.11.016
  [hep-th/0409100].

\bibitem{deAzcarraga:1989mza}
  J.~A.~de Azcarraga, J.~P.~Gauntlett, J.~M.~Izquierdo and P.~K.~Townsend,
  ``Topological Extensions of the Supersymmetry Algebra for Extended Objects,''
  Phys.\ Rev.\ Lett.\  {\bf 63} (1989) 2443.
  doi:10.1103/PhysRevLett.63.2443

\bibitem{Sezgin:1996cj}
  E.~Sezgin,
  ``The M algebra,''
  Phys.\ Lett.\ B {\bf 392} (1997) 323
  doi:10.1016/S0370-2693(96)01576-6
  [hep-th/9609086].

\bibitem{Townsend:1997wg}
  P.~K.~Townsend,
  ``M theory from its superalgebra,''
  In *Cargese 1997, Strings, branes and dualities* 141-177
  [hep-th/9712004].

\bibitem{Hassaine:2003vq}
  M.~Hassaine, R.~Troncoso and J.~Zanelli,
  ``Poincare invariant gravity with local supersymmetry as a gauge theory for the M-algebra,''
  Phys.\ Lett.\ B {\bf 596} (2004) 132
  doi:10.1016/j.physletb.2004.06.067
  [hep-th/0306258].

\bibitem{Hassaine:2004pp}
  M.~Hassaine, R.~Troncoso and J.~Zanelli,
  ``11D supergravity as a gauge theory for the M-algebra,''
  PoS WC {\bf 2004} (2005) 006
  [hep-th/0503220].

\bibitem{Witten:1978mh}
  E.~Witten and D.~I.~Olive,
  ``Supersymmetry Algebras That Include Topological Charges,''
  Phys.\ Lett.\ B {\bf 78} (1978) 97.
  doi:10.1016/0370-2693(78)90357-X

\bibitem{Achucarro:1987nc}
  A.~Achucarro, J.~M.~Evans, P.~K.~Townsend and D.~L.~Wiltshire,
  ``Super p-Branes,''
  Phys.\ Lett.\ B {\bf 198} (1987) 441.
  doi:10.1016/0370-2693(87)90896-3

\bibitem{Schwarz:1995jq}
  J.~H.~Schwarz,
  ``The power of M theory,''
  Phys.\ Lett.\ B {\bf 367} (1996) 97
  doi:10.1016/0370-2693(95)01429-2
  [hep-th/9510086].

\bibitem{Duff:1996aw}
  M.~J.~Duff,
  ``M theory (The Theory formerly known as strings),''
  Int.\ J.\ Mod.\ Phys.\ A {\bf 11} (1996) 5623
   [Subnucl.\ Ser.\  {\bf 34} (1997) 324]
   [Nucl.\ Phys.\ Proc.\ Suppl.\  {\bf 52} (1997) no.1-2,  314]
  doi:10.1142/S0217751X96002583
  [hep-th/9608117].

\bibitem{Townsend:1996xj}
  P.~K.~Townsend,
  ``Four lectures on M theory,''
  In *Trieste 1996, High energy physics and cosmology* 385-438
  [hep-th/9612121].

\bibitem{Bergshoeff:1987cm}
  E.~Bergshoeff, E.~Sezgin and P.~K.~Townsend,
  ``Supermembranes and Eleven-Dimensional Supergravity,''
  Phys.\ Lett.\ B {\bf 189} (1987) 75.
  doi:10.1016/0370-2693(87)91272-X

\bibitem{Duff:1987bx}
  M.~J.~Duff, P.~S.~Howe, T.~Inami and K.~S.~Stelle,
  ``Superstrings in D=10 from Supermembranes in D=11,''
  Phys.\ Lett.\ B {\bf 191} (1987) 70.
  doi:10.1016/0370-2693(87)91323-2

\bibitem{Bergshoeff:1987qx}
  E.~Bergshoeff, E.~Sezgin and P.~K.~Townsend,
  ``Properties of the Eleven-Dimensional Super Membrane Theory,''
  Annals Phys.\  {\bf 185} (1988) 330.
  doi:10.1016/0003-4916(88)90050-4

\bibitem{Townsend:1995kk}
  P.~K.~Townsend,
  ``The eleven-dimensional supermembrane revisited,''
  Phys.\ Lett.\ B {\bf 350} (1995) 184
  doi:10.1016/0370-2693(95)00397-4
  [hep-th/9501068].

\bibitem{Townsend:1995gp}
  P.~K.~Townsend,
  ``P-brane democracy,''
  In *Duff, M.J. (ed.): The world in eleven dimensions* 375-389
  [hep-th/9507048].

\bibitem{Castellani:1985ka}
  L.~Castellani, A.~Ceresole, S.~Ferrara, R.~D'Auria, P.~Fre and E.~Maina,
  ``The Complete $N=3$ Matter Coupled Supergravity,''
  Nucl.\ Phys.\ B {\bf 268} (1986) 317.

\bibitem{DAuria:2000afl}
  R.~D'Auria, S.~Ferrara and S.~Vaula,
  ``Matter coupled F(4) supergravity and the AdS(6) / CFT(5) correspondence,''
  JHEP {\bf 0010} (2000) 013
  doi:10.1088/1126-6708/2000/10/013

\bibitem{Andrianopoli:1996ve}
  L.~Andrianopoli, R.~D'Auria and S.~Ferrara,
  ``U duality and central charges in various dimensions revisited,''
  Int.\ J.\ Mod.\ Phys.\ A {\bf 13} (1998) 431
  doi:10.1142/S0217751X98000196

\bibitem{DallAgata:2003sjo}
  G.~Dall'Agata, R.~D'Auria, L.~Sommovigo and S.~Vaula,
  ``D = 4, N=2 gauged supergravity in the presence of tensor multiplets,''
  Nucl.\ Phys.\ B {\bf 682} (2004) 243
  doi:10.1016/j.nuclphysb.2004.01.014
  [hep-th/0312210].

\bibitem{DAuria:2004yjt}
  R.~D'Auria, L.~Sommovigo and S.~Vaula,
  ``N = 2 supergravity Lagrangian coupled to tensor multiplets with electric and magnetic fluxes,''
  JHEP {\bf 0411} (2004) 028
  doi:10.1088/1126-6708/2004/11/028
  [hep-th/0409097].

\bibitem{Andrianopoli:2011zj}
  L.~Andrianopoli, R.~D'Auria, L.~Sommovigo and M.~Trigiante,
  ``D=4, N=2 Gauged Supergravity coupled to Vector-Tensor Multiplets,''
  Nucl.\ Phys.\ B {\bf 851} (2011) 1
  doi:10.1016/j.nuclphysb.2011.05.007
  [arXiv:1103.4813 [hep-th]].

\bibitem{Andrianopoli:2014aqa}
  L.~Andrianopoli and R.~D'Auria,
  ``N=1 and N=2 pure supergravities on a manifold with boundary,''
  JHEP {\bf 1408} (2014) 012
  doi:10.1007/JHEP08(2014)012
  [arXiv:1405.2010 [hep-th]].

\bibitem{Andrianopoli:2018ymh}
  L.~Andrianopoli, B.~L.~Cerchiai, R.~D.'Auria and M.~Trigiante,
  ``Unconventional supersymmetry at the boundary of AdS$_{4}$ supergravity,''
  JHEP {\bf 1804} (2018) 007
  doi:10.1007/JHEP04(2018)007
  [arXiv:1801.08081 [hep-th]].

\bibitem{DAuria:1987tdr}
  R.~D'Auria, P.~Fr\'{e}, M.~Raciti and F.~Riva,
  ``Anomaly Free Supergravity in $D=10$. 1. The Bianchi Identities and the Bosonic Lagrangian,''
  Int.\ J.\ Mod.\ Phys.\ A {\bf 3} (1988) 953.
  doi:10.1142/S0217751X88000436

\bibitem{Ceresole:1995jg}
  A.~Ceresole, R.~D'Auria, S.~Ferrara and A.~Van Proeyen,
  ``Duality transformations in supersymmetric Yang-Mills theories coupled to supergravity,''
  Nucl.\ Phys.\ B {\bf 444} (1995) 92
  doi:10.1016/0550-3213(95)00175-R
  [hep-th/9502072].

\bibitem{Ceresole:1992su}
  A.~Ceresole, R.~D'Auria, S.~Ferrara, W.~Lerche and J.~Louis,
  ``Picard-Fuchs equations and special geometry,''
  Int.\ J.\ Mod.\ Phys.\ A {\bf 8} (1993) 79
  doi:10.1142/S0217751X93000047
  [hep-th/9204035].

\bibitem{Ceresole:1993qq}
  A.~Ceresole, R.~D'Auria, S.~Ferrara, W.~Lerche, J.~Louis and T.~Regge,
  ``Picard-Fuchs equations, special geometry and target space duality,''
  AMS/IP Stud.\ Adv.\ Math.\  {\bf 1} (1996) 281.

\bibitem{Billo:1995ge}
  M.~Bill\`{o}, A.~Ceresole, R.~D'Auria, S.~Ferrara, P.~Fr\'{e}, T.~Regge, P.~Soriani and A.~Van Proeyen,
  ``A Search for nonperturbative dualities of local N=2 Yang-Mills theories from Calabi-Yau threefolds,''
  Class.\ Quant.\ Grav.\  {\bf 13} (1996) 831
  doi:10.1088/0264-9381/13/5/007
  [hep-th/9506075].

\bibitem{Ceresole:1995ca}
  A.~Ceresole, R.~D'Auria and S.~Ferrara,
  ``The Symplectic structure of N=2 supergravity and its central extension,''
  Nucl.\ Phys.\ Proc.\ Suppl.\  {\bf 46} (1996) 67
  doi:10.1016/0920-5632(96)00008-4
  [hep-th/9509160].

\bibitem{Ceresole:1993nz}
  A.~Ceresole, R.~D'Auria and T.~Regge,
  ``Duality group for Calabi-Yau 2 moduli space,''
  Nucl.\ Phys.\ B {\bf 414} (1994) 517
  doi:10.1016/0550-3213(94)90439-1
  [hep-th/9307151].

\bibitem{Cadavid:1995bk}
  A.~C.~Cadavid, A.~Ceresole, R.~D'Auria and S.~Ferrara,
  ``Eleven-dimensional supergravity compactified on Calabi-Yau threefolds,''
  Phys.\ Lett.\ B {\bf 357} (1995) 76
  doi:10.1016/0370-2693(95)00891-N
  [hep-th/9506144].

\bibitem{DAuria:1981zjr}
  R.~D'Auria, P.~Fr\'{e} and A.~J.~da Silva,
  ``Geometric Structure of $N=1 D=10$ and $N=4 D=4$ Superyang-mills Theory,''
  Nucl.\ Phys.\ B {\bf 196} (1982) 205.
  doi:10.1016/0550-3213(82)90036-0

\bibitem{DAuria:1981ddz}
  R.~D'Auria and T.~Regge,
  ``Gravity Theories With Asymptotically Flat Instantons,''
  Nucl.\ Phys.\ B {\bf 195} (1982) 308.
  doi:10.1016/0550-3213(82)90402-3

\bibitem{Chandia:1997hu}
  O.~Chandia and J.~Zanelli,
  ``Topological invariants, instantons and chiral anomaly on spaces with torsion,''
  Phys.\ Rev.\ D {\bf 55}, 7580 (1997)
  doi:10.1103/PhysRevD.55.7580
  [hep-th/9702025].

\end{thebibliography}
\end{document}